\documentclass[12pt]{article}
\usepackage{graphicx}
\usepackage{epstopdf}
\usepackage{mathrsfs}
\usepackage{amssymb}
\usepackage{color}
\usepackage{amsmath}
\usepackage{ascmac}
\usepackage{amsthm}
\usepackage{booktabs}
\usepackage{lscape}
\usepackage{dcolumn}
\usepackage{longtable}
\usepackage{dcolumn}
\usepackage{arydshln}
\usepackage{bm}
\usepackage{url}
\usepackage{caption}
\usepackage{subfig}
\usepackage{setspace}
\usepackage{cases}
\usepackage{comment}
\usepackage{multirow}

\newcommand{\eg}{e.g.}
\newcommand{\etal}{et al.}
\newcommand{\ie}{i.e.}

\newcommand{\vx}{\mathbf{x}}

\newcommand{\eps}{\epsilon}
\newcommand{\vbeta}{\bm{\beta}}

\makeatletter
  \renewenvironment{thebibliography}[1]{%

    \section*{\refname}\@mkboth{\refname}{\refname}%
    \list{\@biblabel{\@arabic\c@enumiv}}%
          {\settowidth\labelwidth{\@biblabel{#1}}%
          \leftmargin\labelwidth
          \advance\leftmargin\labelsep
          \setlength\baselineskip{10truept}
          \setlength\itemsep{-2truept}
          \@openbib@code
          \usecounter{enumiv}%
          \let\p@enumiv\@empty
          \renewcommand\theenumiv{\@arabic\c@enumiv}}%
    \sloppy
    \clubpenalty4000
    \@clubpenalty\clubpenalty
    \widowpenalty4000%
    \sfcode`\.\@m}
    {\def\@noitemerr
      {\@latex@warning{Empty `thebibliography' environment}}%
    \endlist}
\makeatother

\begin{document}

\title{Chain effects of clean water: The Mills--Reincke phenomenon in early twentieth-century Japan
}

\author{
Tatsuki Inoue
\thanks{Corresponding author. Graduate School of Economics, The University of Tokyo, 7-3-1, Hongo, Bunkyo-ku, Tokyo
113-0033, Japan. E-mail: inoue-tatsuki245@g.ecc.u-tokyo.ac.jp. Telephone number: +81-3-5841-5543.}
~and 
Kota Ogasawara
\thanks{Department of Industrial Engineering and Economics, Tokyo Institute of Technology, 2-12-1, Ookayama, Meguro-ku, Tokyo 263-8522, Japan (E-mail: ogasawara.k.ab@m.titech.ac.jp).}
}

\date{}
\maketitle
\begin{abstract}

This study explores the validity of chain effects of clean water, which are known as the ``Mills--Reincke phenomenon,'' in early twentieth-century Japan.
Recent studies have reported that water purifications systems are responsible for huge contributions to human capital.
Although some studies have investigated the instantaneous effects of water-supply systems in pre-war Japan, little is known about the chain effects of these systems.
By analyzing city-level cause-specific mortality data from 1922--1940, we find that a decline in typhoid deaths by one per 1,000 people decreased the risk of death due to non-waterborne diseases such as tuberculosis and pneumonia by 0.742--2.942 per 1,000 people.
Our finding suggests that the observed Mills--Reincke phenomenon could have resulted in the relatively rapid decline in the mortality rate in early twentieth-century Japan.

\bigskip

\bigskip


\end{abstract}

\newpage
\section{Introduction} \label{sec:intro}

Waterborne diseases caused considerable losses of human capital throughout the twentieth century (Preston and van de Walle 1978; Evans 1987; Troesken 2004). 
Previous studies have therefore regarded the implementation of water purification systems as being responsible for the huge improvements in public health.
The pioneering and influential study by Cutler and Miller (2005) showed that the development and application of water purification technologies was responsible for roughly 40\% of the decline in the mortality rate from 1900 to 1940.
Subsequent studies also found the similar improving effects of clean water both in European and in Asian countries (Macassa \etal~2006; Jaadla and Puur 2016; Ogasawara and Matsushita 2018; Peltola and Saaritsa 2019).\footnote{Alsan and Goldin (2019) and Kesztenbaum and Rosenthal (2017) showed the importance of sewers in reducing the mortality rate in Massachusetts and Paris, respectively.}
However, the recent study by Anderson \etal~(2018) found more moderate impacts of water purification technology on crude and infant mortality rates than those shown by Cutler and Miller (2005).
Brown and Guinnane (2018) also provided contradictory views on these improving effects of safe water in Bavaria between 1825 and 1910.

The present study seeks to contribute to this debate by adding new evidence on the improving effects of modern water-supply systems on the historical decline in the mortality rate.
While a large number of previous studies have shown the instantaneous effects of clean water on mortality rates, we try to show the chain (i.e., persistent) effects of safe water on mortality rates by testing the Mills--Reincke phenomenon.
The Mills--Reincke phenomenon is an epidemiological proposition arguing that the purification of polluted water could not only reduce deaths due to typhoid in the initial stage, but also those due to other infectious diseases in the later stage (Sedgwick and MacNutt 1910; Evans 1987).
Epidemiological case studies have provided evidence of this phenomenon in large cities in France and the United States (Preston and van de Walle 1978; Crimmins and Condran 1983).
Economic studies of this phenomenon are more scarce; however, the work by Ferrie and Troesken (2008) showed that for every death from typhoid fever prevented by water purification, three or more deaths from other causes were also prevented in Chicago from 1850 to 1925.
While these studies have predominantly focused on the cases of a few large cities, this study is the first to use a more comprehensive dataset covering multiple cities in prewar Japan.

We find that eliminating typhoid fever infection decreased the risk of non-waterborne diseases. 
Our estimates show that for each additional death due to typhoid, there were approximately two deaths due to tuberculosis and pneumonia; this magnitude is indeed greater than that observed previously in Chicago (Ferrie and Troesken 2008).
The composition of the cause-specific deaths suggests that the national malady (\textit{kokuminby\=o}), especially tuberculosis, was more likely to be improved by the eradication of typhoid fever.
This finding not only supports the evidence provided by Ferrie and Troesken but also adds more comprehensive evidence on the chain effects of clean water in industrializing Japan.

This study contributes to the broader literature in the following two ways.
First, we complement the abovementioned discussion on the role of water purification technology in mitigating historical mortality declines (Anderson \etal~2018; Cutler and Miller 2019).
The results of this study indeed support the evidence provided by previous studies of the important role of clean-water technology (\eg, Alsan and Goldin 2019).
Second, we explore the validity of the Mills--Reincke phenomenon using comprehensive city-level mortality data from the early twentieth century.
Expanding Ferrie and Troesken's (2008) strategy to capture the mechanisms responsible for the appearance of the phenomenon, we find that the chain effects of clean water were still considered to be substantial.\footnote{Our cost/benefit analysis in \ref{app:cba} also indicates that the contributions of clean-water technologies to historical declines in mortality have been sufficiently large to exceed the costs of installation, as demonstrated by Cutler and Miller (2005) and Ferrie and Troesken (2008).}
This study highlights the importance of investigating not only the instantaneous effects but also the subsequent chain effects of clean water when we discuss the impacts of the implementation of water purification systems on historical mortality declines.

The rest of the paper proceeds as follows:
Section~\ref{sec:back} provides an overview of mortality rate trends in pre-war Japan, and also illustrates the features of the Mills--Reincke phenomenon.
Section~\ref{sec:data} describes the data that we used.
Section~\ref{sec:mills} explains our empirical strategy and presents the main results.
Section~\ref{sec:con} concludes.
\section{Background} \label{sec:back}
\subsection{Waterworks and mortality rates in pre-war Japan} \label{sec:21}

\begin{figure}[]
    \centering
    \subfloat[Tap water use and typhoid death rates] {\label{fig:ts_wa_tf}\includegraphics[width=0.5\textwidth]{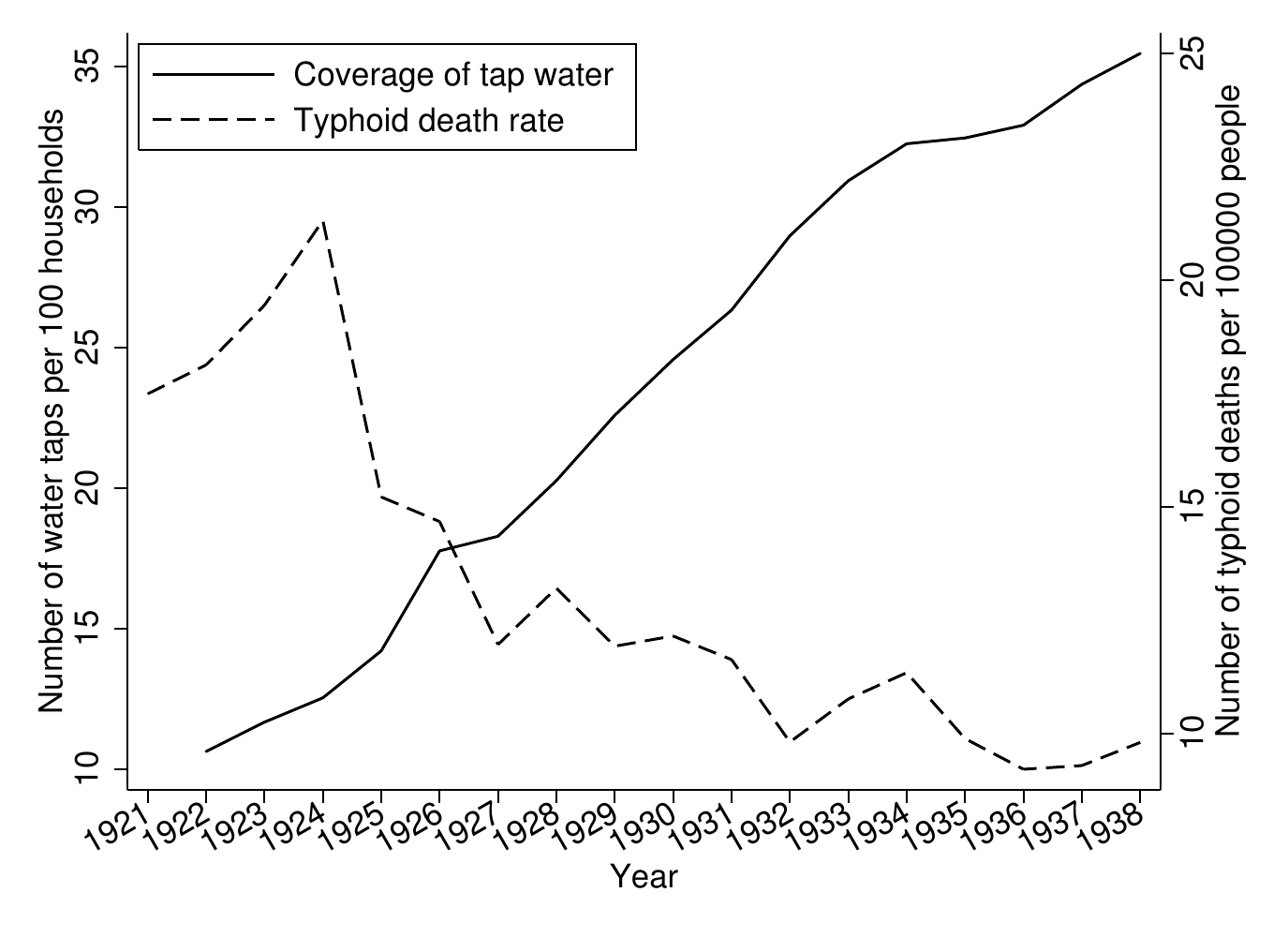}}
    \subfloat[Death rates in major cities]{\label{fig:ts_death}\includegraphics[width=0.5\textwidth]{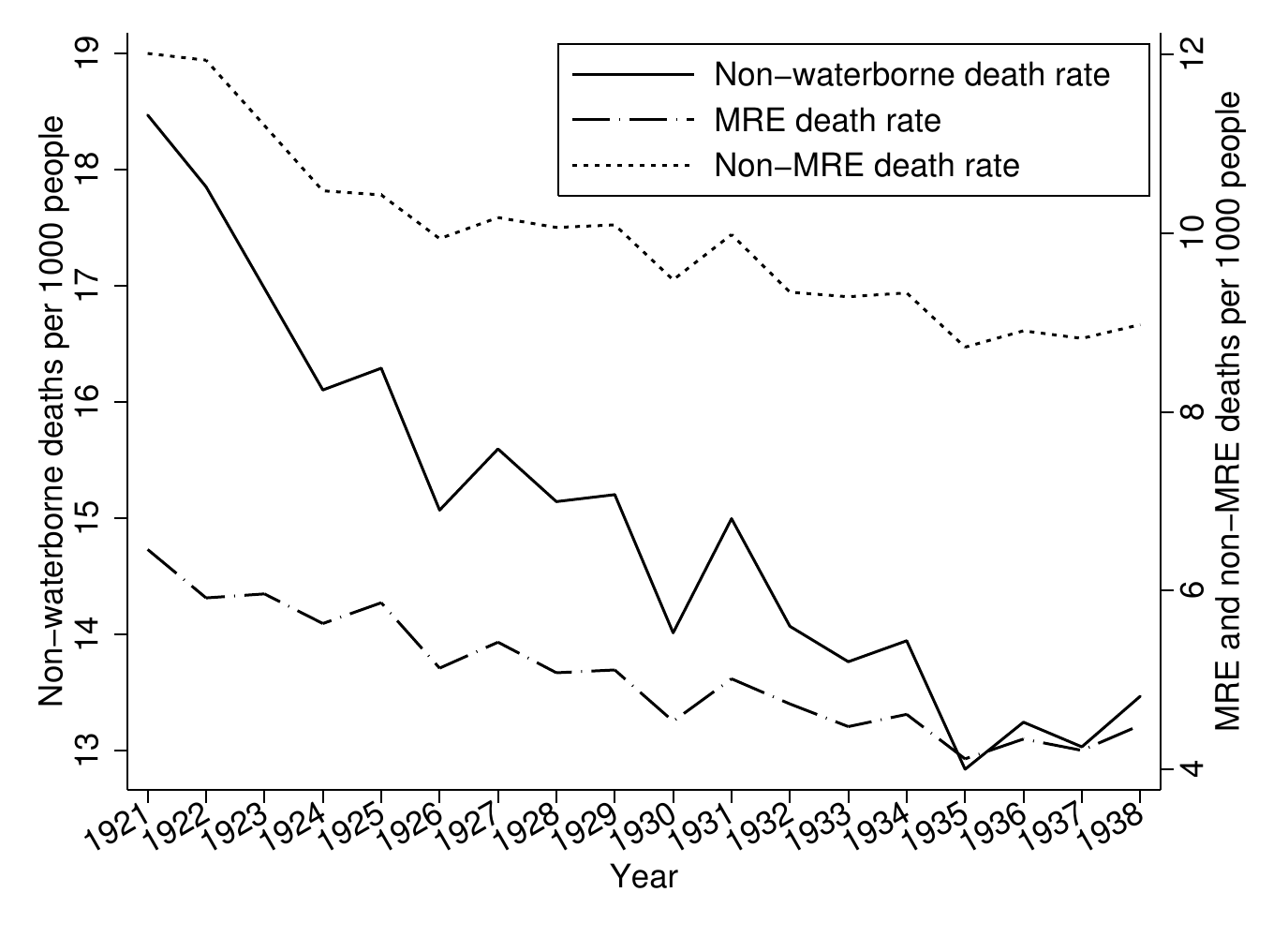}}
    \caption{Tap water coverage and cause-specific death rates}
    \label{fig:ts_wa_tf_death}
    \scriptsize{\begin{minipage}{450pt}
    Notes:
    Tap water coverage is defined as the average number of water taps per 100 households in cities.
    Non-waterborne death is defined as death from causes other than typhoid fever, paratyphoid, dysentery, cholera, and diarrhea.
    The definitions of MRE and non-MRE death rates are provided in the text.
    Data are shown for cities with populations of more than 100,000.
    We excluded data from Tokyo city in 1932 because limited information is available on cause-specific deaths in this year, as a consequence of merging municipalities.
    We also excluded data on non-waterborne and non-MRE death rates in 1923, to eliminate the influence of the Great Kant\=o Earthquake.
    Sources:
    Japan Water Works Association (1967, p.200) and Statistics Bureau of the Cabinet (1924a--1932a; 1934a--1942a; 1924b--1939b). suidoutoukei
    \end{minipage}}
\end{figure}

In Japan, modern waterworks were first constructed in Yokohama city in 1887.
Subsequently, the port cities of Hakodate and Nagasaki introduced water-supply systems in 1888.
Modern waterworks, however, were installed only in the most highly populated cities and open ports at the beginning because the scope of the governmental subsidy was limited (Japan Water Works Association 1967).
The Waterworks Ordinance enacted in 1890 required the construction cost of waterworks to be covered by public funds and thus a subsidy from the national government was necessary for municipalities to introduce modern water-supply systems.
Therefore, installations and expansions of modern waterworks only quickly spread after 1918 and 1920 when the government twice expanded the scope of the subsidy.
Indeed, modern systems were installed in only seven locations in 1900, rising to 55 in 1921 and 345 in 1940 (Japan Water Works Association 1967).
Likewise, Fig.~\ref{fig:ts_wa_tf} shows that the number of water taps per 100 households in cities increased from 10.63\% in 1922 to 35.45\% in 1938.

The most important difference between old and modern systems is whether they had water purification technologies.
Old waterworks simply ran water from rivers and springs to urban areas without any clarification facilities.
Moreover, their wooden pipes had decayed over a long time (Japan Water Works Association 1967).
Thus, the water supplied through the old waterworks was incredibly polluted and carried pathogens.
For instance, the University of Tokyo and Sanitary Bureau of the Home Department conducted a survey of water quality in Tokyo around 1880 and concluded that drinking water in urban areas was dirty like ``thinned urine'' (Bureau of Waterworks, Tokyo Metropolitan Government 1999, pp.6--7).
By contrast, modern waterworks used cast iron pipes and water purified by filtration (and chlorination in some cities) in accordance with water quality standards that are still applicable today (Japan Water Works Association 1967).
These improvements in water-supply systems made tap water sufficiently clean to prevent waterborne infections (Ogasawara and Matsushita 2018).

Figure~\ref{fig:ts_wa_tf} illustrates the death rates from typhoid fever, a waterborne disease, in 1921--1938 as well as the coverage of tap water.
The typhoid death rate indeed decreased rapidly as the use of water-supply systems spread.
However, it should be noted that a downward trend in mortality rates was also observed for deaths due to non-waterborne disease as shown in Fig.~\ref{fig:ts_death}.
This means that improvements in both waterborne and non-waterborne risks of death contributed to the declines in mortality, which invites the question, ``Were water-supply systems responsible for the reductions in non-waterborne mortality?''
If one considers the Mills--Reincke phenomenon, then the mortality transitions that were observed in Japan provide interesting new evidence.

Figure~\ref{fig:ts_death} shows rates of Milles--Reincke effect-associated (MRE) deaths, which are defined as deaths from tuberculosis, pneumonia, bronchitis, meningitis, heart disease.
These causes of death are likely to be subject to the Mills--Reincke effect as discussed in detail in Subsection~\ref{sec:sec22}. Rates of death are reported per 1,000 people and are for major cities.
The MRE death rate dropped from 6.46 in 1921 to 4.49 in 1938, which is a 30.50\% reduction.
On the other hand, Fig.~\ref{fig:ts_death} also shows rates of non-MRE deaths, which are defined as deaths from causes other than the MRE and waterborne deaths.
The rates of non-MRE deaths in 1921 and 1938 were 12.01 and 8.98 per 1,000 people, respectively, which amounts to a 25.23\% reduction.
Clearly, the MRE death rate declined more rapidly than the non-MRE death rate.
This divergence strongly suggests that the Mills--Reincke phenomenon was observed in Japan, as previously reported for European countries.
Therefore, to understand the entirety of the effects that improving water quality had on mortality, we investigated the chain of effects in terms of the Mills--Reincke phenomenon.

\subsection{Externalities of typhoid fever} \label{sec:sec22}

Typhoid fever, which is an oral infectious disease, is caused by the bacterium \textit{Salmonella enterica} serotype Typhi (\textit{S. Typhi}) (Huang and DuPont 2005).
The onset of symptoms is marked by fever and malaise.
Patients often have a fever, a dull frontal headache, nausea, a dry cough, a coated tongue, splenomegaly, relative bradycardia, constipation, diarrhea, and any other symptoms, but have few physical signs (Parry \etal~2002, p.1774).
Typhoid fever often does not threaten human life directly and the case-fatality rate is reported to be approximately 10--20\% (World Health Organization 2011, p.1).
This relatively low case-fatality rate likely understates its importance.
The most important feature of typhoid fever is that it usually causes non-typhoid diseases or worsens chronic diseases.
Therefore, it has often been observed that, as water quality improved in Western nations during the late nineteenth and early twentieth centuries, there were reductions in both typhoid and non-typhoid death rates. 
The negative correlation between the purification of polluted water supplies and non-typhoid death rates is called the \textit{Mills--Reincke phenomenon} because it was discovered independently by Hiram F. Mills and J. J. Reincke in 1893--1894.%
\footnote{See Appendix~\ref{sec:seca1} for finer details of the typhoid fever and a brief overview of the epidemiological studies of this phenomenon that were performed in the early twentieth century.}

The main factor that could cause the Mills--Reincke phenomenon is considered to be complications from typhoid fever.
Numerous extra-intestinal complications can occur with \textit{S. Typhi} infection, including the involvement of the central nervous, cardiovascular, pulmonary, bone and joints, hepatobiliary, and genitourinary systems (Huang and DuPont 2005).
To confirm that deaths from complications accounted for some proportion of deaths that were not due to typhoid fever itself, we compiled a unique survey report, in which the frequencies of deaths from complications of typhoid fever were recorded.
In this survey, complications were investigated for $1,214$ patients diagnosed with typhoid fever and hospitalized in Komagome Hospital in Tokyo during 1932--1933 (Tokyo City Office 1935).
According to this survey, $96.21$\% (\ie, $1,168$) of patients had complications during the survey period.%
\footnote{Table~\ref{tab:casecomp1} in Appendix~\ref{sec:seca_comp} shows the numbers of complication patients by disease category.}
Table~\ref{tab:komagome} summarizes the frequencies of deaths due to typhoid fever or complications among typhoid patients.
Of the $1,214$ patients with typhoid fever, $92$ ($7.58$\%) had deaths attributed to the complications, implying that if patients with typhoid fever were affected with other diseases, approximately $7.6$\% of them might have died of these complications.
Moreover, since the total number of deaths among the patients was reportedly $172$, the number of deaths due to complications per 100 total deaths is estimated to be $53.49$\% ($92$/$172$).
This implies that more than half of the deaths in patients with typhoid fever were ultimately due to non-typhoid diseases.

Among the complications, the shares of deaths from enterorrhagia and perforation (which are considered to be direct consequences of typhoid fever) were relatively high ($20.35$\% and $5.81$\%, respectively).
The more important fact is that pneumonia, beriberi, and meningitis also accounted for relatively large proportions of the deaths ($13.95$\%, $6.4$\%, and $3.49$\% of the $172$ typhoid-related deaths, respectively).
In addition, mumps and bronchitis were also listed as complications that had fatal consequences. 
However, they accounted for only $1.16$\% and $1.74$\% of the total deaths, respectively.
The above findings suggest that improvements in the typhoid death rate are more likely to reduce the death rates from mumps, bronchitis, pneumonia, beriberi, and meningitis, as well as digestive diseases such as enterorrhagia and perforation.

The abovementioned findings imply that various complications were derived from infections with typhoid fever.
These complications caused \textit{additional} deaths from non-typhoid causes, such as pulmonary and cardiovascular system diseases, as well as digestive system diseases.

\begin{table}[!t]

\begin{center}
\caption{The number of cause-specific deaths among typhoid patients}
\label{tab:komagome}
\scriptsize
\scalebox{0.87}[1]{
\begin{tabular}{lrrr}
\toprule
Causes of death	&Number of deaths&Proportion to $1,214$ patients of typhoid fever (\%)&Share of $172$ total deaths (\%)\\\hline

Total&172&14.17&100\\
Typhoid fever&80&6.59&46.51\\
Complications								&92	&7.58	&53.49\\
\hspace{10pt}Acute hepatic insufficiency		&1	&0.08	&0.58\\
\hspace{10pt}Enterorrhagia				&35	&2.88	&20.35\\
\hspace{10pt}Perforation					&10	&0.82	&5.81\\
\hspace{10pt}Mumps					&2	&0.16	&1.16\\
\hspace{10pt}Bronchitis					&3	&0.25	&1.74\\
\hspace{10pt}Pneumonia					&24	&1.98	&13.95\\
\hspace{10pt}Beriberi					&11	&0.91	&6.40\\
\hspace{10pt}Meningitis					&6	&0.49	&3.49\\
\bottomrule
\end{tabular}
}
{\scriptsize
\begin{minipage}{155mm}
Notes:
For each complication, the rate of deaths from the complication is the number of deaths due to the complication per 100 patients with typhoid fever.
In total, the cohort included 1,214 patients with typhoid fever.
The share of total deaths is the number of deaths due to complications per 100 deaths from any cause.
Source: Tokyo City Office (1935, pp.1--9).
\end{minipage}
}

\end{center}
\end{table}

\section{Data} \label{sec:data}

We hypothesized that the improvement of water-supply systems reduced both waterborne and non-waterborne deaths by preventing typhoid fever.
To confirm this hypothesis by empirical analyses, the present study used different panel datasets for our two main dependent variables: the non-typhoid death rate and the MRE death rate.
The characteristics of the study samples are as follows:
To estimate the effects on the non-typhoid death rate, we compiled panel data from 108 cities pertaining to the period between 1922 and 1940, including almost all of the cities interspersed across the whole of Japan.
In fact, approximately 92.38\% of the city-dwelling Japanese citizens were included in our study cohort during the study period.
Regarding the MRE death rate, the number of deaths from each cause is not described for every city.
Nevertheless, we could compile panel data on cause-specific deaths in 26 cities with populations greater than 100,000 between 1922 and 1936.%
\footnote{Cities with populations greater than 50,000 are included for 1922.}


In our hypothesis, typhoid fever was the key factor contributing to the Mills--Reincke phenomenon.
We use the death rate from typhoid fever as an independent variable that indexes typhoid epidemic and severity level.
The typhoid death rate (\textit{Typhoid}) is defined as the number of deaths from typhoid fever per 1,000 people.
In addition, we use the typhoid incidence rate (\textit{TIR}) as an alternative measure of epidemic level. 
This is defined as the number of cases of typhoid fever per 1,000 people.
However, the measurement error of typhoid cases was larger than that of typhoid deaths.%
\footnote{For example, some people who were affected by typhoid fever masked their disease (Doi 1925).}
Thus, we take the death rate as primary measure in our estimates.
The data on the number of typhoid deaths and cases are obtained from statistical reports: \textit{Nihonteikoku shiin tokei}, \textit{Josuido tokei}, \textit{Eiseikyoku nenpo}, and \textit{Eisei nenpo}.


We use the following measures of mortality rates as the dependent variables.
The non-typhoid death rate (\textit{Non-typhoid}) is defined as the number of deaths from causes other than typhoid fever per 1,000 people.
Since this measure was also used in Ferrie and Troesken (2008), we are able to compare the results of the present study with their previous research.
However, the non-typhoid death rate includes deaths that could not possibly be considered to have been affected by typhoid, such as drowning and freezing deaths.
This might increase the gap between the underlying truth and our interpretation of estimated results.

Therefore, we also use the MRE death rate (\textit{MRE death}).
We define MRE deaths as deaths from tuberculosis, pneumonia, bronchitis, meningitis, and heart disease based on medical and historical evidence.
These diseases involve the central nervous, cardiovascular, and pulmonary systems, which can occur with \textit{S. Typhi} infection (Huang and DuPont 2005).%
\footnote{We categorize tuberculosis, which was known as one of the two major national maladies in Japan, as MRE deaths; it was not listed as showing complications by Huang and DuPont (2005), as tuberculosis is rare in modern developed countries because of the development of effective drugs (Zumla \etal~2013).}
In addition, they were frequently observed as complications of typhoid fever in pre-war Japan.
If our hypothesis is correct, the relationship between the MRE death rate and the typhoid death rate should be significantly positive.

Furthermore, to investigate the Mills--Reincke phenomenon in more detail, we estimate effects on cause-specific death rates.
As a placebo test, our main estimates also use the death rate for causes that were not categorized as MRE death.
We expect that preventing typhoid fever would contribute to decreases in deaths that met the definition of MRE death, but not affect other deaths, namely those due to scarlet fever, smallpox, diphtheria, whooping cough, measles, influenza, beriberi, and nephritis.
These tracheal and renal system diseases and nutritional deficiency disease are not considered to be complications of typhoid fever in medical studies (see Huang and DuPont 2005).
Our main sources for the data on the cause-specific deaths are \textit{Nihonteikoku shiin tokei} and \textit{Shiin tokei}.
The materials are based on official statistics after the first national census conducted in 1920 and the statistics are known to be sufficiently accurate (see Ito 1987).


We included control variables representing demographic, socioeconomic, and meteorological characteristics.
The demographic variables included the number of citizens, shares of various age groups, and the sex ratios of the age groups.
For socioeconomic variables, we included the size of the financial budget per capita, the number of doctors per 100 people, and the proportion of industrial workers in the population. 
For meteorological variables, we included the annual mean temperature, annual mean humidity, and annual mean actual sunshine duration.
These variables could control for spatial heterogeneity in the risk of infectious diseases (Ni \etal~2014; World Health Organization 2009).
In addition, to control for the the consequences of the Great Kant\=o Earthquake, we included an indicator variable that takes the value one for both Tokyo and Yokohama in 1923.
\ref{app:data} provides the finer details of our data, definitions of the variables, and data sources.

\section{Testing Mills--Reincke phenomenon} \label{sec:mills}
\subsection{Non-typhoid death} \label{sec:non}

To capture the size and impact of the Mills--Reincke phenomenon in early twentieth-century Japan, we employed the following city fixed effects approach as our identification strategy.
Our baseline model is given by
\begin{eqnarray}
\text{\textit{Non-typhoid}}_{it} = \alpha + \delta \text{\textit{Typhoid}}_{it} + \vx'_{it}\vbeta + v_{i} + u_{t} + t\gamma_{i} + e_{it }\label {eq:base}
\end{eqnarray}
where $i$ indexes cities from 1 to 108 and $t$ indexes years from 1922 to 1940.
The variable $\textit{Typhoid}_{it}$ is the typhoid death rate, $\textit{Non--typhoid}_{it}$ is the non-typhoid death rate,
$\vx'_{it}$ is a vector of city characteristics, and $e_{it}$ is a random error term.
$v_{i}$ and $u_{t}$ represent city and year fixed effects, respectively.
$t\gamma_{i}$ indicates a city-specific linear time trend.
The coefficient $\delta$ is our parameter of interest, and its estimate $\hat{\delta}$ measures the impact of typhoid fever on the non-typhoid death rate.

Table~\ref{tab:non} reports our results for the non-typhoid death rate estimated by Eq.~(\ref{eq:base}).
In all specifications, we controlled for the consequences of the Great Kant\=o Earthquake.
The result in columns (3) and (7) report the estimates from our baseline specification, in which we controlled for all characteristics.
Evidently, the estimates are stable across the different specifications and different measures.
In columns (1)--(4), all estimated coefficients of \textit{Typhoid} are significantly positive.
This result suggests that typhoid fever caused complications which led to death.
Moreover, the coefficient of $\textit{Non-typhoid}_{t-1}$, which is insignificant in column (4), implies that typhoid deaths were not affected by other death previous year.

To understand the contributions of the present study, we should compare our results with the previous literature.
Columns (1)--(4) show that one additional typhoid death per 1,000 people lead to increases in deaths from other causes by 0.994, 1.013, 1.033, and 1.092 per 1,000 people, respectively.
On the other hand, in the case of Chicago from 1850 to 1925, the feasible estimate of the Mills--Reincke effect was between 4 and 7 (Ferrie and Troesken 2008, p.13).
Our estimates are obviously much smaller than those presented in the previous study.
A possible explanation of this difference is competing risks.
If someone died from typhoid fever, the cause of death was understandably typhoid fever rather than any other disease.
Thus, an increase in typhoid deaths among typhoid cases would potentially decrease the number of deaths due to complications.
In pre-war Japan, the case-fatality rate of typhoid fever was roughly 20\%, whereas the rate was 5--10\% in the United States in 1850--1925 (Ferrie and Troesken 2008, p.7).%
\footnote{Typhoid case-fatality rate is generally approximately 10--20\% (World Health Organization 2011, p.1),}
This higher case-fatality rate might lower the Mills--Reincke effect in Japan in comparison with the United States.

In columns (5)--(8), the estimated coefficients are 0.170, 0.172, 0.175, and 0.206, respectively.
Clearly, all of the results remained significantly positive, suggesting that our main finding is not sensitive to the definition of the key variable that was used to measure typhoid epidemic level.
Moreover, it is noteworthy that the result for incidence rate imply a mechanism for the Mills--Reincke phenomenon.
The estimated coefficient in column (7) means that for one additional typhoid-infected person there were 0.175 additional non-typhoid deaths. 
Further, since the average typhoid case-fatality rate in our sample is 19.13\%, the result in column (3) suggests that one additional \textit{incidence} case of typhoid fever per 1,000 people increased non-typhoid death rates by 0.198‰ (\ie, 0.1913 $\times$ 1.033).
Therefore, the impacts of the increase in the typhoid death rate were likely to be larger than the impacts of the increase in the incidence rate.
This implies that complications were caused by serious cases of typhoid, which are  reflected by the death rate, rather than by mild cases.
In other words, it is possible that relatively more non-typhoid deaths were prevented by decreases in the more serious forms of typhoid.

\begin{table}[!t]

\begin{center}
\caption{Effects of typhoid fever on non-typhoid death}
\label{tab:non}
\scriptsize

\begin{tabular*}{160mm}{lcccccccc}
\toprule

&(1)&(2)&(3)&(4)&(5)&(6)&(7)&(8)\\
\hline

\textit{Typhoid}
&0.994***&1.013***&1.033***&1.092***&&&&\\
&(0.330)&(0.326)&(0.318)&(0.337)&&&&\\

\textit{TIR}
&&&&&0.170***&0.172***&0.175***&0.206***\\
&&&&&(0.063)&(0.061)&(0.060)&(0.063)\\

\textit{Non-typhoid}$_{t-1}$
&&&&$-$0.035&&&&$-$0.056*\\
&&&&(0.035)&&&&(0.032)\\

\hline

Control variables
&&&&\\
\hspace{8pt}Demographic controls
&Yes&Yes&Yes&Yes&Yes&Yes&Yes&Yes\\
\hspace{8pt}Socioeconomic controls
&No&Yes&Yes&Yes&No&Yes&Yes&Yes\\
\hspace{8pt}Meteorological controls
&No&No&Yes&Yes&No&No&Yes&Yes\\

Fixed effects and trends
&Yes&Yes&Yes&Yes&Yes&Yes&Yes&Yes\\

$R$-squared
&0.7024&0.7032&0.7053&0.7124&0.7048&0.7052&0.7072&0.7159\\

Number of clusters
&108&108&108&107&108&108&108&107\\
Number of observations
&1,464&1,457&1,457&1,290&1,444&1,437&1,437&1,270\\

\bottomrule\end{tabular*}

{\scriptsize
\begin{minipage}{155mm}
Notes: 
Observations are at the city-year level.
The dependent variable is the non-typhoid death rate.
All regressions include the indicator variable for the Great Kant\=o Earthquake.
Details of the data sources of each variable used in the regression are provided in the text and the \ref{app:data}.
***, **, and * represent statistical significance at the 1\%, 5\%, and 10\% levels, respectively.
Cluster-robust standard errors are in parentheses.
\end{minipage}
}

\end{center}
\end{table}

\subsection{Cause-specific death} \label{sec:cause}

Since the non-typhoid death rate includes some noise such as non-MRE deaths, we investigate the effects related to the Mills--Reincke phenomenon in detail using cause-specific death rates.
The specification of our fixed effects model is given by
\begin{eqnarray}
\text{\textit{y}}_{it} = \alpha + \delta \text{\textit{Typhoid}}_{it} + \vx'_{it}\vbeta + v_{i} + u_{t} + t\gamma_{i} + e_{it}\label {eq:cause}
\end{eqnarray}
where $i$ indexes cities from 1 to 26 and $t$ indexes years from 1922 to 1936.
The variable $y_{it}$ is the dependent variable, such as the MRE death rate or the cause-specific death rate.
The other variables are defined as in Eq.~(\ref{eq:base}).

We further employed the fixed effects two-stage least squares (FE-2SLS) approach using the number of water taps per 100 households (\textit{Water}) as an instrumental variable to address a possible omitted variable problem when controlling for fixed effects.%
\footnote{
Although the modern sewage system is also an effective public health facility (Alsan and Goldin 2019), the coverage of sewage systems among citizens continued to be low in interwar Japan; for instance, the share of households with flushable toilets in 1935 was only 0.16\% (Ogasawara and Matsushita 2019). Since the traditional system of circulating human waste from urban to rural areas was well developed before the introduction of chemical fertilizers, urban households had little need to be connected to the sewage system for waste disposal for a long time. Maeda (2008 pp. 67--70) suggested that this feature of traditional disposal systems led to delays in the installation of a flushing toilet.}
Our specification of FE-2SLS model is then given by
%
\begin{eqnarray}
\text{\textit{Typhoid}}_{it} &=& \alpha_{1} + \zeta\text{\textit{Water}}_{it} + \vx'_{it}\bm{\lambda} + \nu_{i} + \mu_{t} + t\rho_{i} + \eps_{it}  \label{eq:fir}\\
\text{\textit{y}}_{it} &=& \alpha_{2} + \delta \widehat{\text{\textit{Typhoid}}}_{it} + \vx'_{it}\vbeta + v_{i} + u_{t} + t\gamma_{i} + e_{it}  \label{eq:sec}
\end{eqnarray}
%
where $i$ indexes cities from 1 to 26 and $t$ indexes years from 1922 to 1936.
As defined in Eq.~(\ref{eq:base}), the variable $\vx'_{it}$ is a vector of city characteristics, $\nu_{i}$ ($v_{i}$) and $\mu_{i}$ ($u_{t}$) represent city and year fixed effects, and $t\rho_{i}$ ($t\gamma_{i}$) indicates a city-specific linear time trend.
In Eq.~(\ref{eq:fir}) of the first stage,  $\textit{Typhoid}_{it}$ is the typhoid death rate, $\textit{Water}_{it}$ is the tap water coverage, and $\eps_{it}$ is a random error term.
In Eq.~(\ref{eq:sec}) of the second stage, we used the fitted value $\widehat{\text{\textit{Typhoid}}}_{it}$ instead of $\textit{Typhoid}_{it}$ to identify the effect of typhoid fever instrumented by $\textit{Water}_{it}$.
The variable $\textit{y}_{it}$ is the dependent variable: the MRE death rate and each of the cause-specific death rates.%
\footnote{The \textit{Non-typhoid} variable is likely to correlate with the popularization of tap water because it includes deaths due to waterborne diseases, such as dysentery and diarrhea.
Therefore, we employed the FE-2SLS model for only the MRE and the cause-specific death rates.
}
$e_{it}$ is a random error term.

The instrument \textit{Water} is considered to satisfy the exclusion restriction for the following reasons.
First, both the timing of installation and the transition of tap water coverage were exogenous owing to the characteristics of the natural environment and unpredictable events.
Installations of modern water-supply systems were influenced by geographical characteristics such as a warm climate, light rainfall, and shallow rivers; natural disasters such as droughts, floods, and wildfires; political issues such as water rights, residents' opposition, delayed land acquisitions, and conflict between city councils; and increases in the import value of cast iron pipes.
In addition, the expansion of the number of water taps was influenced by changes in the natural state of water sources, outbreak of war, and consolidation of municipalities as well as the factors described above.
\footnote{See Ogasawara and Matsushita (2018) for finer details of the uncertainties of modern water-supply projects in early twentieth-century Japan.}
Therefore, we can exploit the exogenous variation over time using tap water coverage rather than the timing of the installation of waterworks.

Second, there was no direct relationship between tap water coverage and non-waterborne death rates.
To confirm that modern water-supply systems did not directly decrease non-waterborne diseases, we estimate the effects of tap water coverage on death rates from non-waterborne diseases, namely tuberculosis, pneumonia, bronchitis, meningitis, heart disease, scarlet fever, smallpox, diphtheria, whooping cough, measles, influenza, beriberi, and nephritis.
Our specification is similar to that in Eq.~(\ref{eq:cause}) but we use \textit{Water} instead of \textit{Typhoid} as the key independent variable.
All the demographic, socioeconomic, and meteorological variables, which are important factors in deaths because of typhoid fever as well as other diseases, are included as control variables.
Table~\ref{tab:water} reports the estimation results.
The finding that the estimated coefficients of \textit{Water} are insignificant for all death rates regardless of MRE or non-MRE deaths indicates that waterworks had no direct effects.
\begin{table}[!t]

\begin{center}
\caption{Effects of tap water coverage on cause-specific deaths}
\label{tab:water}
\scriptsize

\begin{tabular*}{160mm}{l@{\extracolsep{\fill}}D{.}{.}{3}cccc}
\toprule

Dependent variable&\multicolumn{1}{c}{\textit{Water}}&\shortstack{\\All control\\variables}&\shortstack{\\Fixed effects\\and trends}&\shortstack{\\Number of\\clusters}&\shortstack{\\Number of\\observations}\\
\hline

MRE death\\
\hspace{10pt}\textit{MRE death}
&-0.0237&\multirow{2}{*}{Yes}&\multirow{2}{*}{Yes}&\multirow{2}{*}{26}&\multirow{2}{*}{258}\\
&(0.0147)&&&&\\

\hspace{10pt}Tuberculosis
&-0.0056&\multirow{2}{*}{Yes}&\multirow{2}{*}{Yes}&\multirow{2}{*}{26}&\multirow{2}{*}{258}\\
&(0.0043)&&&&\\

\hspace{10pt}Pneumonia
&-0.0099&\multirow{2}{*}{Yes}&\multirow{2}{*}{Yes}&\multirow{2}{*}{26}&\multirow{2}{*}{258}\\
&(0.0063)&&&&\\

\hspace{10pt}Bronchitis
&-0.0022&\multirow{2}{*}{Yes}&\multirow{2}{*}{Yes}&\multirow{2}{*}{26}&\multirow{2}{*}{258}\\
&(0.0016)&&&&\\

\hspace{10pt}Meningitis
&-0.0055&\multirow{2}{*}{Yes}&\multirow{2}{*}{Yes}&\multirow{2}{*}{26}&\multirow{2}{*}{258}\\
&(0.0036)&&&&\\

\hspace{10pt}Heart disease
&-0.0005&\multirow{2}{*}{Yes}&\multirow{2}{*}{Yes}&\multirow{2}{*}{26}&\multirow{2}{*}{258}\\
&(0.0015)&&&&\\
\\

Non-MRE death\\
\hspace{10pt}\textit{Non-MRE death}
&-0.0019&\multirow{2}{*}{Yes}&\multirow{2}{*}{Yes}&\multirow{2}{*}{26}&\multirow{2}{*}{258}\\
&(0.0054)&&&&\\

\hspace{10pt}Scarlet fever
&-0.0001&\multirow{2}{*}{Yes}&\multirow{2}{*}{Yes}&\hspace{5truept}\multirow{2}{*}{91}&\hspace{-6truept}\multirow{2}{*}{1,120}\\
&(0.0001)&&&&\\

\hspace{10pt}Smallpox
&-0.0002&\multirow{2}{*}{Yes}&\multirow{2}{*}{Yes}&\hspace{5truept}\multirow{2}{*}{36}&\multirow{2}{*}{520}\\
&(0.0002)&&&&\\

\hspace{10pt}Diphtheria
&-0.0004&\multirow{2}{*}{Yes}&\multirow{2}{*}{Yes}&\multirow{2}{*}{106}&\hspace{-6truept}\multirow{2}{*}{1,262}\\
&(0.0004)&&&&\\

\hspace{10pt}Whooping cough
&-0.0007&\multirow{2}{*}{Yes}&\multirow{2}{*}{Yes}&\multirow{2}{*}{26}&\multirow{2}{*}{258}\\
&(0.0008)&&&&\\

\hspace{10pt}Measles
&0.0007&\multirow{2}{*}{Yes}&\multirow{2}{*}{Yes}&\multirow{2}{*}{26}&\multirow{2}{*}{258}\\
&(0.0023)&&&&\\

\hspace{10pt}Influenza
&0.0002&\multirow{2}{*}{Yes}&\multirow{2}{*}{Yes}&\multirow{2}{*}{26}&\multirow{2}{*}{258}\\
&(0.0005)&&&&\\

\hspace{10pt}Beriberi
&0.0007&\multirow{2}{*}{Yes}&\multirow{2}{*}{Yes}&\multirow{2}{*}{26}&\multirow{2}{*}{258}\\
&(0.0021)&&&&\\

\hspace{10pt}Nephritis
&-0.0011&\multirow{2}{*}{Yes}&\multirow{2}{*}{Yes}&\multirow{2}{*}{26}&\multirow{2}{*}{258}\\
&(0.0017)&&&&\\

\bottomrule
\end{tabular*}

{\scriptsize
\begin{minipage}{155mm}
Notes: 
Observations are at the city-year level.
The third column presents the estimated coefficient of the independent variable \textit{Water} and its standard error for each dependent variable.
All control variables are the same as the estimate reported in Table~\ref{tab:cause}. 
All regressions include the indicator variable for the Great Kant\=o Earthquake.
Details of the data sources of each variable used in the regression are provided in the \ref{app:data}.
***, **, and * represent statistical significance at the 1\%, 5\%, and 10\% levels, respectively.
Cluster-robust standard errors are in parentheses. 
\end{minipage}
}

\end{center}
\end{table}

Third, modern waterworks improved water quality but did not dramatically change the time taken to access water, which could be used for something else to improve health if it was not needed.
Although the ground water was occasionally contaminated, Tokyo city had 187,169 wells, that is, 3.27 per 100 square meters in 1939 (Bureau of Waterworks, Tokyo Metropolitan Government 1999, p.193; Tokyo Institute for Municipal Research 1939, p.14).
In addition, many cities had old waterworks that drew water from rivers and springs to residential areas without any purification systems before the installation of modern waterworks (Japan Water Works Association 1967).
These facts indicate that access to water would have been easy in Japan even if no modern waterworks had been installed.
These evidences support the validity of our instrument if we control for the appropriate city characteristics.

\begin{table}[!t]

\begin{center}
\caption{Effects of typhoid fever on cause-specific deaths, FE and FE-2SLS}
\label{tab:cause}
\scriptsize

\begin{tabular*}{160mm}{l@{\extracolsep{\fill}}D{.}{.}{5}D{.}{.}{5}cccc}
\toprule
Panel A: Results of FE and FE-2SLS
&&&&&&\\

&\multicolumn{2}{c}{Typhoid death rate}&\multirow{2}{*}{\shortstack{\\All control\\variables}}&\multirow{2}{*}{\shortstack{\\Fixed effects\\and trends}}&\multirow{2}{*}{\shortstack{\\Number of\\clusters}}&\multirow{2}{*}{\shortstack{\\Number of\\observations}}\\
\cmidrule(rl){2-3}
Dependent variable&\multicolumn{1}{c}{FE}&\multicolumn{1}{c}{FE-2SLS}&&&\\
\hline

MRE death\\
\hspace{10pt}\textit{MRE death}
&0.742\text{**}&2.942\text{***}&Yes&Yes&26&258\\
&(0.280)&(1.038)&&&&\\

\hspace{10pt}Tuberculosis
&0.113&0.696\text{**} &Yes&Yes&26&258\\
&(0.086)&(0.336)&&&&\\

\hspace{10pt}Pneumonia
&0.464\text{**}&1.233\text{**}&Yes&Yes&26&258\\
&(0.202)&(0.553)&&&&\\

\hspace{10pt}Bronchitis
&0.065 &0.275\text{*}&Yes&Yes&26&258\\
&(0.055)&(0.152)&&&&\\

\hspace{10pt}Meningitis
&0.063&0.682\text{**}&Yes&Yes&26&258\\
&(0.058)&(0.270)&&&&\\

\hspace{10pt}Heart disease
&0.038 &0.057&Yes&Yes&26&258\\
&(0.047)&(0.115)&&&&\\
\\

Non-MRE death\\
\hspace{10pt}\textit{Non-MRE death}
&0.162&0.242&Yes&Yes&26&258\\
&(0.161)&(0.542)&&&&\\

\hspace{10pt}Scarlet fever
&0.009\text{**}&0.091&Yes&Yes&91&\hspace{-6truept}1,120\\
&(0.004)&(0.079)&&&&\\

\hspace{10pt}Smallpox
&-0.002&0.071&Yes&Yes&36&520\\
&(0.009)&(0.052)&&&&\\

\hspace{10pt}Diphtheria
&0.012&0.291&Yes&Yes&\hspace{-4truept}106&\hspace{-6truept}1,262\\
&(0.012)&(0.225)&&&&\\

\hspace{10pt}Whooping cough
&0.013&0.083&Yes&Yes&26&258\\
&(0.035)&(0.155)&&&&\\

\hspace{10pt}Measles
&0.011&-0.092&Yes&Yes&26&258\\
&(0.119)&(0.323)&&&&\\

\hspace{10pt}Influenza
&-0.032&-0.022&Yes&Yes&26&258\\
&(0.020)&(0.081)&&&&\\

\hspace{10pt}Beriberi
&0.092\text{*}&-0.082&Yes&Yes&26&258\\
&(0.053)&(0.194)&&&&\\

\hspace{10pt}Nephritis
&0.080&0.143&Yes&Yes&26&258\\
&(0.052)&(0.163)&&&&\\

\end{tabular*}

\begin{tabular*}{160mm}{l@{\extracolsep{\fill}}cccc}
\hline
\\
Panel B: First stage results of FE-2SLS
&&&&\\
&(1)&(2)&(3)&(4)\\
\hline

\textit{Water}
&$-$0.0080***&$-$0.0014**&$-$0.0034**&$-$0.0014**\\
&(0.0024)&(0.0007)&(0.0013)&(0.0006)\\
\hline

All control variables
&Yes&Yes&Yes&Yes\\
City- and Year- fixed effects
&Yes&Yes&Yes&Yes\\
Linear time trends
&Yes&Yes&Yes&Yes\\
Kleibergen-Paap \textit{rk} Wald $F$-statistic
&11.59&4.48&6.52&4.60\\
$R$-squared
&0.671&0.560&0.583&0.540\\
Number of clusters
&26&91&36&106\\
Number of observations
&258&1,120&520&1,262\\

\bottomrule
\end{tabular*}

{\scriptsize
\begin{minipage}{155mm}
Notes: 
Observations are at the city-year level.
All regressions include the indicator variable for the Great Kant\=o Earthquake.
Details of the data sources of each variable used in the regression are provided in the text and the \ref{app:data}.
***, **, and * represent statistical significance at the 1\%, 5\%, and 10\% levels, respectively.
Cluster-robust standard errors are in parentheses. 
\end{minipage}
}

\end{center}
\end{table}

Table~\ref{tab:cause} reports the results for MRE deaths and non-MRE deaths in Panel~A.%
\footnote{Robustness checks and cost-benefit analysis are reported in \ref{app:rob} and D.
}
The results of FE were estimated by our baseline specification, as shown in column (3) of Table~\ref{tab:non}.
The estimated coefficient for the MRE death rate by fixed effects model is positive and statistically significant, which is consistent with the medical and historical evidences.
The coefficient is 0.742 and thus implies that, for one additional typhoid death per 1,000 people, there were 0.742 additional MRE deaths per 1,000 people.
Since this value is approximately equal to three-quarters of the coefficient reported in column (3) of Table~\ref{tab:non}, it is likely that the impact on MRE deaths accounted for a large proportion of the impact on non-typhoid deaths, although we should take into account the difference in samples between the analyses.

Our FE results show that while there is no significant relationship between the typhoid death rate and death rates of tuberculosis, bronchitis, meningitis, and heart disease, there is evidence that the incidence of typhoid increased the number of deaths from pneumonia (significant at the 5\% level), even when only pneumonia was considered.
The coefficient for pneumonia suggests that one additional typhoid death per 1,000 people caused 0.464 pneumonia deaths per 1,000 people.
In contrast to MRE deaths, the FE results indicate that typhoid had no significant effect on non-MRE deaths, except for scarlet fever and beriberi.
The results correspond with our hypothesis that the relationships between typhoid and non-MRE deaths are negligible.

Panel~A of Table~\ref{tab:cause} also presents our second stage results of the FE-2SLS model.
The specification includes the same control variables as FE model. 
Obviously, the estimated coefficient for the MRE death rate is positive and significant, which is consistent with the fixed effects estimate.
This result suggests that modern water-supply systems reduced MRE deaths by preventing typhoid fever.
Our point estimate means that for one additional typhoid death per 1,000 people, there were 2.942 MRE deaths per 1,000 people.
The effect in the FE-2SLS regression is approximately four times larger than that in the FE regression.

The coefficients for the tuberculosis, pneumonia, bronchitis, and meningitis death rates are also significantly positive, whereas the coefficients for all the cause-specific death rates categorized as non-MRE deaths are insignificant.
This result provides the evidence to support our hypothesis.
In contrast to the results that were obtained from the fixed effects model, the estimated coefficients for beriberi and scarlet fever are not significant.
A possible interpretation of this result is that these diseases were directly affected by clean water.

In contrast to the comparison with the result for the non-typhoid death rate, our estimated MRE effect on respiratory deaths (i.e., tuberculosis, pneumonia, and bronchitis deaths) via the FE-2SLS model is twice as great as the effect estimated by Ferrie and Troesken (2008, p.11).
While they estimated the MRE effect on respiratory deaths in Chicago to be roughly one, our estimate is roughly two.
This result is consistent with the rapid decline in the mortality rate in Japanese cities relative to US cities.
One possible explanation of this difference is that tuberculosis was more prevalent in pre-war Japan than in Western countries (Johnston 1995; Hunter 2003), suggesting that tuberculosis patients were the potential beneficiary of clean water at that time.

Panel~B of Table~\ref{tab:cause} shows the first-stage results of the FE-2SLS model.
All the estimated coefficients of \textit{Water} are significantly negative regardless of the number of observations.
This result suggests that underidentification due to an irrelevant instrument is less likely to be problematic.
Moreover, the $F$-statistic reported in column (1) is more than 10, indicating that we can reject the weak instrument assumption according to the criteria of Staiger and Stock (1997).
This finding means that most of the second-stage results reported in Panel A are reliable. 
However, we cannot reject the weak instrument assumption in columns (2)--(4) of Panel B, implying that the second-stage estimates for scarlet fever, smallpox, and diphtheria shown in Panel A may be biased and should be regarded as the upper bound of the effects.

\section{Conclusion}\label{sec:con}

This study focuses on the potential contributions of amelioration in water quality on health: the effects of the Mills--Reincke phenomenon.
We investigate these effects using panel data from Japan during the 1920s and 1930s and our estimates are consistent with the results reported by the related studies described in the Introduction.

Our findings are as follows.
First, the Mills--Reincke phenomenon was observed in Japanese cities.
In addition, historical records show the many complications of typhoid fever, supporting our hypothesis that typhoid fever was the key factor behind the Mills--Reincke phenomenon.
Second, our estimates from the fixed effects models and FE-2SLS models suggest that a decline in typhoid deaths by one per 1,000 people decreased MRE deaths (\ie, tuberculosis, pneumonia, bronchitis, meningitis, and heart disease death rates) by 0.742 and 2.942 per 1,000 people, respectively.
We also find that this chain effect on respiratory disease was greater than that observed in the United States.

From a broader view, the results of this study provide evidence to support the efficacy of public health improvements such as modern waterworks.
In 2000, 216,510 people were estimated to have died of typhoid fever, especially in developing Asian countries (Crump \etal~2004, p.346; Siddiqui \etal~2006).
Therefore, our estimate suggests that if public water purification systems eradicate typhoid fever, an additional 160,650--636,972 people could escape death induced by typhoid.

\section*{Acknowledgements}
We wish to thank the participants in the seminars at the Tokyo Institute of Technology for their helpful comments on the paper. 
The work was supported by the fund for JSPS Research Fellow (Grant Number: 17J03825) and JSPS KAKENHI (Grant Number: 17K03096). 
There are no conflicts of interest to declare.
All errors are our own.

\clearpage
\thispagestyle{empty}

\begin{center}
\qquad

\qquad

\qquad

\qquad

\qquad

\qquad

{\LARGE \textbf{Online Appendices\\
(Supplemental materials for review)
}}
\end{center}

\clearpage

\appendix
\def\thesection{Appendix~\Alph{section}}
\def\thesubsection{\Alph{section}.\arabic{subsection}}

\setcounter{page}{1}

\section{Background appendix} \label{app: typhoid}
\setcounter{table}{0} \renewcommand{\thetable}{A.\arabic{table}}
\setcounter{table}{0} \renewcommand{\thetable}{A.\arabic{table}}

\subsection{Typhoid fever and Mills--Reincke phenomenon} \label{sec:seca1}

\textit{Salmonella enterica} serotype Typhi (\textit{S. Typhi}), which causes typhoid fever, are invasive bacteria that rapidly and efficiently pass though the intestinal mucosa of humans to reach the reticuloendothelial system (World Health Organization 2011, p.7).%
\footnote{Paratyphoid fever is caused by \textit{Salmonella enterica} serotype Paratyphi (\textit{S. Paratyphi}) A, B, and C.
In most endemic areas, typhoid fever accounts for 75--80\% of enteric fever cases (World Health Organization 2011, p.1).
In the case of pre-war Japan, paratyphoid fevers were less likely to be observed, as we note later.}
The incubation period is usually relatively long, and it takes 8--14 days until the presentation of clinical typhoid fever (World Health Organization 2011, p.8).
Patients often have fever, influenza-like symptoms with chills, a dull frontal headache, malaise, anorexia, nausea, poorly localized abdominal discomfort, a dry cough, and myalgia, but have few physical signs.%
\footnote{The symptoms described here are based on the report of Parry \etal~(2002, p.1774).}
The other common symptoms are a coated tongue, tender abdomen, hepatomegaly, splenomegaly, relative bradycardia, constipation, and diarrhea.
Although the fever is low grade initially, it rises progressively; the fever often reaches 39 to 40$^\circ$C by the second week and is sustained.
In 5--30\% of cases, a few rose spots of approximately 2 to 4 mm in diameter occur on the abdomen and chest.

Today, it is known that fluoroquinolones (a kind of antibiotics) are effective for the treatment of typhoid fever (Parry \etal~2002, p.1775).
In regard to vaccination, the first report of randomized controlled trial of typhoid vaccine was published in 1962 (Engels \etal~1998, p.110).
However, it was not possible to depend on antibiotics or widespread, effective vaccination in Japan during the 1920s and 1930s.

The Mills--Reincke phenomenon was observed in Europe in the end of nineteenth century for the first time.
Hiram F.Mills, a member of the State Board of Health of Massachusetts, found that not only the typhoid deaths but also the non-typhoid deaths had declined shortly after the introduction of filtered and purified water supply in to Lawrence, Massachusetts in September 1893.
At the same time, J.J. Reincke, a health officer of the city of Hamburg in Germany, also observed the same phenomenon in that city other introducing filtered public water-supply system in May 1893 (Sedgwick and MacNutt 1910).
Subsequently, both Allen Hazen, a famous civil and Sanitary engineer, and W.T.Sedgwick and J.Scott MacNutt, a professor of Biology at the Massachusetts Institute of Technology and a health officer of Orange in New Jersey, observed this Mills--Reincke phenomenon in a lot of cities in the United States as well as European countries (Sedgwick and MacNutt 1910; Hazen 1914).

Although Fink (1917) suspected the Mills--Reincke phenomenon and suggested the difficulties to determine the exact relationship between the reduction in typhoid deaths and the decline in non-typhoid deaths, the other subsequent studies also found the evidence of the Mills--Reincke phenomenon at that time (\eg, McGee 1920).
These epidemiological study in the early twentieth century found that the non-typhoid death rates such as tuberculosis, pneumonia, bronchitis, malaria, heart disease, marasmus as well as the other gastrointestinal diseases had declined in line with the improvement of water quality (see Sedgwick and MacNutt 1910; Fink 1917).

As described in the Introduction, a recent study by Ferrie and Troesken (2008) also revealed that 35--56\% percent of the reduction in the crude death rate in the city of Chicago between 1850 and 1925 was related to water purification, the associated eradication of typhoid fever, and related Mills--Reincke effects.
They found that water purification had reduced not only the death rates from waterborne diseases, such as typhoid fever and diarrhea, but also the death rates from non-typhoid deaths, such as influenza, pneumonia, heart disease, and tuberculosis.
They commented that, ``the typical typhoid survivor was so weakened and compromised by the disease that he or she would later succumb to some other infectious disease like tuberculosis, or die of kidney or hear failure'' (Ferrie and Troesken 2008, p.15).

\subsection{Complications from typhoid fever} \label{sec:seca_comp}

\begin{table}[!h]

\def\arraystretch{0.9}
\begin{center}
\caption{The number of patients affected by complications, by disease category}
\label{tab:casecomp1}
\scriptsize

\begin{tabular}{lrr}
\toprule
Diseases category&Number of patients&Proportion to $1,214$ patients of typhoid fever (\%)\\\hline

Total							&1,168	&96.21\\
Skin diseases					&126		&10.38\\
Digestive diseases				&181		&14.91\\
Periproctitis					&5		&0.41\\
Otorhinolaryngologic diseases		&287		&23.64\\
Respiratory diseases				&313		&25.78\\
Cardiovascular diseases			&82		&6.75\\
Genitourinary diseases 			&103		&8.48\\
Eye diseases					&7		&0.58\\
Neuropathy					&33		&2.72\\
Surgical diseases				&4		&0.33\\
Other complications				&27		&2.22\\
(Reinfection)					&3		&0.25\\
\bottomrule

\end{tabular}

{\scriptsize
\begin{minipage}{390pt}
Notes:
This table shows the number of patients affected by complications in a cohort of $1,214$ patients with typhoid fever.
Skin diseases include herpes, hives, bedsore, furunculosis, abscess, jaundice, and subcutaneous bleeding.
Digestive diseases include acute hepatic insufficiency, stomach cramps, omphalitis, enterorrhagia, perforation, and intestinal catarrh.
Periproctitis includes anal prolapse, anal fistula, and marasmus.
Otorhinolaryngologic diseases include pharyngodynia, pyorrheal gums, stomatitis and pulpitis, mumps, glue ear, nosebleed, and hearing loss.
Respiratory diseases include bronchitis, pneumonia, pleurisy, tuberculosis, hilar lymphadenitis, and bronchial asthma.
Cardiovascular diseases include beriberi, venous thrombosis, valvular disease, and phrenocardia.
Genitourinary diseases include urogenital disease, nephritis, nephrosis, renal tuberculosis, pyelitis, cystitis, anuresis, orchitis, endometritis, Bartholin's abscess, and conceive.
Eye diseases include visual disturbance, conjunctivitis, phlyctenular conjunctivitis, keratitis parenchymatosa, and cataract.
Neuropathy include meningism, cerebral hemorrhage, and hemiplegia.
Surgical diseases include periostitis, mastitis, and panaritium.
Other complications include struma, thrush, dysentery, whooping cough, roundworm, measles, and mercurial eruption.
Reinfection is the number of patients with reinfection due to typhoid fever.
The disease category is based on Tokyo City Office (1935), and thus some diseases we classify as a digestive or respiratory disease in main text are included in a different category.
Source: Tokyo City Office (1935, pp.1--9).
\end{minipage}
}

\end{center}
\end{table}

Table~\ref{tab:casecomp1} shows the numbers of patients affected by complications by disease category.
According to this survey, $1,168$ of the patients were infected with diseases other than typhoid fever during the times that they had stayed in the hospital.
This figure is quite striking because it means that $96.21$\% (\ie, $1,168$/$1,214$) of the patients had complications during the survey period, implying that infection with typhoid fever led to near twice as cases of disease overall.
Skin, digestive, otorhinolaryngologic, and respiratory diseases were recorded in $126$, $181$, $287$, and $313$ patients, respectively.
Each of these four disease categories was observed to affect more than $10$\% of the patients with typhoid fever.
Altogether, approximately $74.7$\% of the patients with typhoid fever were affected by diseases belonging to at least one of the four categories.

As we described in Subsection 2.2, this survey investigated patients diagnosed with typhoid fever and hospitalized in Komagome Hospital in Tokyo during 1932--1933 (Tokyo City Office 1935, p.1).
It is then suggested that these patients were, at least, not hospitalized due to other diseases. 
However, it is likely that those with other health problems had worse immune systems and had such complication-like conditions before they contracted typhoid.
Therefore, although they were less likely to be infected by the other complications before hospitalization, as the reason for the hospitalization was typhoid fever, we must be careful not to conclude that these figures capture the causal relationship between typhoid infection and complications.

\section{Data appendix}\label{app:data}

\setcounter{table}{0} \renewcommand{\thetable}{B.\arabic{table}}
\setcounter{figure}{0} \renewcommand{\thefigure}{B.\arabic{figure}}

We define the MRE deaths as the deaths from tuberculosis, pneumonia, bronchitis, meningitis, and heart disease and also define the non-MRE deaths as the deaths from scarlet fever, smallpox, diphtheria, whooping cough, measles, influenza, beriberi and nephritis.
The definition of non--MRE deaths is different from the definition in Section~\ref{sec:back}.
All cause-specific death rates (including non-tyohoid, MRE, and non-MRE death rates) are defined as the number of deaths from each disease per 1,000 people.
The data on tuberculosis, pneumonia, meningitis, heart disease, beriberi, bronchitis, whooping cough, measles, influenza, and nephritis are based on \textit{Nihonteikoku shiin tokei} (\textit{Statistics of Causes of Death of the Empire of Japan}; \textit{SCDEJ}) and \textit{Shiin tokei} (\textit{Statistics of Causes of Death}; \textit{SCD}).
While the data on above diseases is described in only \textit{SCDEJ} and \textit{SCD}, the data on scarlet fever, smallpox, and diphtheria---these diseases were categorized as the reportable communicable disease during---is described in \textit{SCDEJ}, \textit{SCD}, \textit{Eiseikyoku nenpo} (\textit{Annual Report of the Sanitary Bureau}; \textit{ARSB}), and \textit{Eisei nenpo} (\textit{Annual Report on Sanitation}; \textit{ARS}).
Hence, we supplemented additionally information about reportable communicable disease.

For demographic variables, we control for the population, the shares of various age groups, and the sex ratio of the age groups.
Each variable is defined as the natural logarithm of total population, the number of people in the 0--14, 15--24, 25--59, and 60+ years age group divided by total population in percentage points, and the number of female divided by the number of male in each age group, respectively.
We obtained these data from \textit{Nihonteikoku jinkodotai tokei} (\textit{Vital Statistics of the Empire of Japan}; \textit{VSEJ}) and \textit{Nihon jinkodotai tokei} (\textit{Vital Statistics of Japan}; \textit{VSJ}).
Although the population censuses of Japan are our sources for the data on the number of people in each age group, the census have been conducted only every five years.
Thus, we linearly interpolate this data for the years between quinquennial censuses following the previous literature (\eg, Cutler and Miller 2005; Greenstone and Hanna 2014).

For socioeconomic variable, we include the size of the financial budget per capita which includes expenses for public health, opportunities for consulting a doctor, and the proportion of industrial workers to the population. 
The definitions of these variables are the natural logarithm of annual total city revenue per capita, the number of doctors per 1,000 people, and the number of factory workers per 100 people.
The annual total city revenue in each city are from \textit{Nihonteikoku tokei nenkan} (Statistical Yearbook of the Japanese Empire; SYJE) (vol.42) and \textit{Chiho zaisei gaiyo} (Abstract of Local Public Finance) (1923--1940) published by the Bureau of Local Affairs of the Home Department (1924--1941).
The number of doctors are from the SYJE (vol.43--55) and \textit{Dainihonteikoku tokei nenkan} (Statistical Yearbook of the Greater Japanese Empire; SYGJE) (vol.56--59) published by the Statistics Bureau of the Cabinet (1924--1941).
The number of industrial workers are from \textit{Kojyo tokei hyo} (Factory Statistics) (1922--1938 edition) and \textit{Kogyo tokei hyo} (Industrial Statistics) (1939--40 editions) published by the Statistical Division of the Department of Agriculture and Commerce (1924--1926), Statistical Division of the Department of the Minister of Commerce and Industry (1927--1939), Research Division of the Department of the Minister of Commerce and Industry (1940--1941), and Research Division of the Bureau of Management and Coordination of the Ministry of Commerce and Industry (1942).
The city revenue is adjusted by using Consumer Price Index at 1930 constant market in Japan from the Bank of Japan (1986).
Among socioeconomic variables, the covariates of doctors and factory workers are measured at prefecture level because of historical data limitation.

\begin{table}[!t]

\def\arraystretch{0.9}
\begin{center}
\caption{Summary statistics}
\label{tab:sum}
\scriptsize

\begin{tabular}{lD{.}{}{3}D{.}{.}{-1}D{.}{.}{-1}D{.}{.}{-1}D{.}{.}{-1}}
\toprule
&\multicolumn{1}{c}{Observations}
&\multicolumn{1}{c}{Mean}
&\multicolumn{1}{c}{Std.Dev.}
&\multicolumn{1}{c}{Min}
&\multicolumn{1}{c}{Max}\\
\hline
Panel (a): Estimates for non-typhoid death&&&&&\\
\textit{Non-typhoid}&1,464&17.83&2.97&10.90&40.22\\
\textit{Typhoid}&1,464&0.19&0.18&0.00&2.08\\
\textit{TIR}&1,444&1.02&0.84&0.00&10.54\\

\\
Panel (b): Estimates for cause-specific death&&&&&\\
\textit{MRE death}&258&5.31&1.03&3.47&8.61\\
\hspace{10pt}\textit{Tuberculosis}&258&1.94&0.40&1.16&3.34\\
\hspace{10pt}\textit{Bronchitis}&258&0.34&0.14&0.13&1.18\\
\hspace{10pt}\textit{Pneumonia}&258&1.78&0.47&0.91&4.03\\
\hspace{10pt}\textit{Meningitis}&258&0.76&0.36&0.22&2.13\\
\hspace{10pt}\textit{Heart disease}&258&0.48&0.17&0.13&1.08\\
\textit{Non-MRE death}&258&2.00&0.58&1.01&4.34\\
\hspace{10pt}\textit{Scarlet fever}&1,120&0.01&0.02&0.00&0.36\\
\hspace{10pt}\textit{Smallpox}&520&0.00&0.03&0.00&0.37\\
\hspace{10pt}\textit{Diphtheria}&1,262&0.07&0.06&0.00&0.52\\
\hspace{10pt}\textit{Whooping cough}&258&0.16&0.10&0.02&0.92\\
\hspace{10pt}\textit{Measles}&258&0.23&0.25&0.01&1.70\\
\hspace{10pt}\textit{Influenza}&258&0.09&0.06&0.00&0.36\\
\hspace{10pt}\textit{Beriberi}&258&0.42&0.27&0.09&1.78\\
\hspace{10pt}\textit{Nephritis}&258&1.01&0.26&0.59&2.04\\

\textit{Typhoid}&258&0.26&0.21&0.03&2.08\\
\textit{Water}&258&34.89&18.55&0.00&72.73\\

\\
Panel (c): Control variables&&&&&\\
\textit{Population}
&1,464&11.45&0.95&10.00&15.73\\
\textit{Share of age 0-14}
&1,464&33.78&2.59&27.21&43.53\\
\textit{Share of age 15-24}
&1,464&22.27&2.79&11.25&33.17\\
\textit{Share of age 25-59}
&1,464&38.13&1.95&33.06&43.27\\
\textit{Sex ratio of age 0-14}
&1,464&0.99&0.05&0.52&1.42\\
\textit{Sex ratio of age 15-24}
&1,464&1.02&0.22&0.04&1.83\\
\textit{Sex ratio of age 25-59}
&1,464&0.96&0.10&0.66&1.29\\
\textit{Sex ratio of age 60+}
&1,464&1.35&0.15&0.71&2.36\\
\textit{Doctor}
&1,464&0.75&0.25&0.06&2.15\\
\textit{Industrial worker}
&1,461&3.47&2.44&0.13&12.47\\
\textit{Financial budget}
&1,460&2.79&0.50&0.48&4.90\\
\textit{Temperature}
&1,464&13.80&2.42&4.28&22.53\\
\textit{Humidity}
&1,464&75.33&3.02&67.50&85.75\\
\textit{Actual sunshine duration}
&1,464&176.20&19.41&121.86&232.45\\

\bottomrule
\end{tabular}

{\scriptsize
\begin{minipage}{385pt}
Notes:
All cause-specific death rates are defined as frequencies per 1,000 people.
Source: See \ref{app:data}.
\end{minipage}
}

\end{center}
\end{table}

For meteorological variables, we use the annual mean temperature, the annual mean humidity, and the annual mean actual sunshine duration.
These data are downloaded from the database of the Japan Meteorological Agency.
Although the meteorological observation stations are located in each city in most cases, we replicate some missing data on the observations at the nearest meteorological observing station.
In the case of Naha city located in Okinawa prefecture, we cannot replicate the missing data using the nearest meteorological observation station because of long distance across the sea.
The data are publicly available and can be downloaded from \url{http://www.data.jma.go.jp/gmd/risk/obsdl/index.php} (accessed 15 August 2018).

The indicator variable that controls consequence of the Great Kant\=o Earthquake scores one if a city is Tokyo or Yokohama and year is 1923.
This disaster struck in 1923, and caused serious damage to the Kant\=o region, especially in Tokyo and Yokohama cities (Earthquake Disaster Prevention and Research Committee 1925, p.34).
The coverage of tap water is defined as the number of metered and communal water taps for households and business per 100 household.
The summary statistics are presented in Table~\ref{tab:sum}.

\begin{figure}[!t]
    \centering
    \subfloat[108 small to large cities] {\label{fig:mapfull}\includegraphics[width=0.35\textwidth]{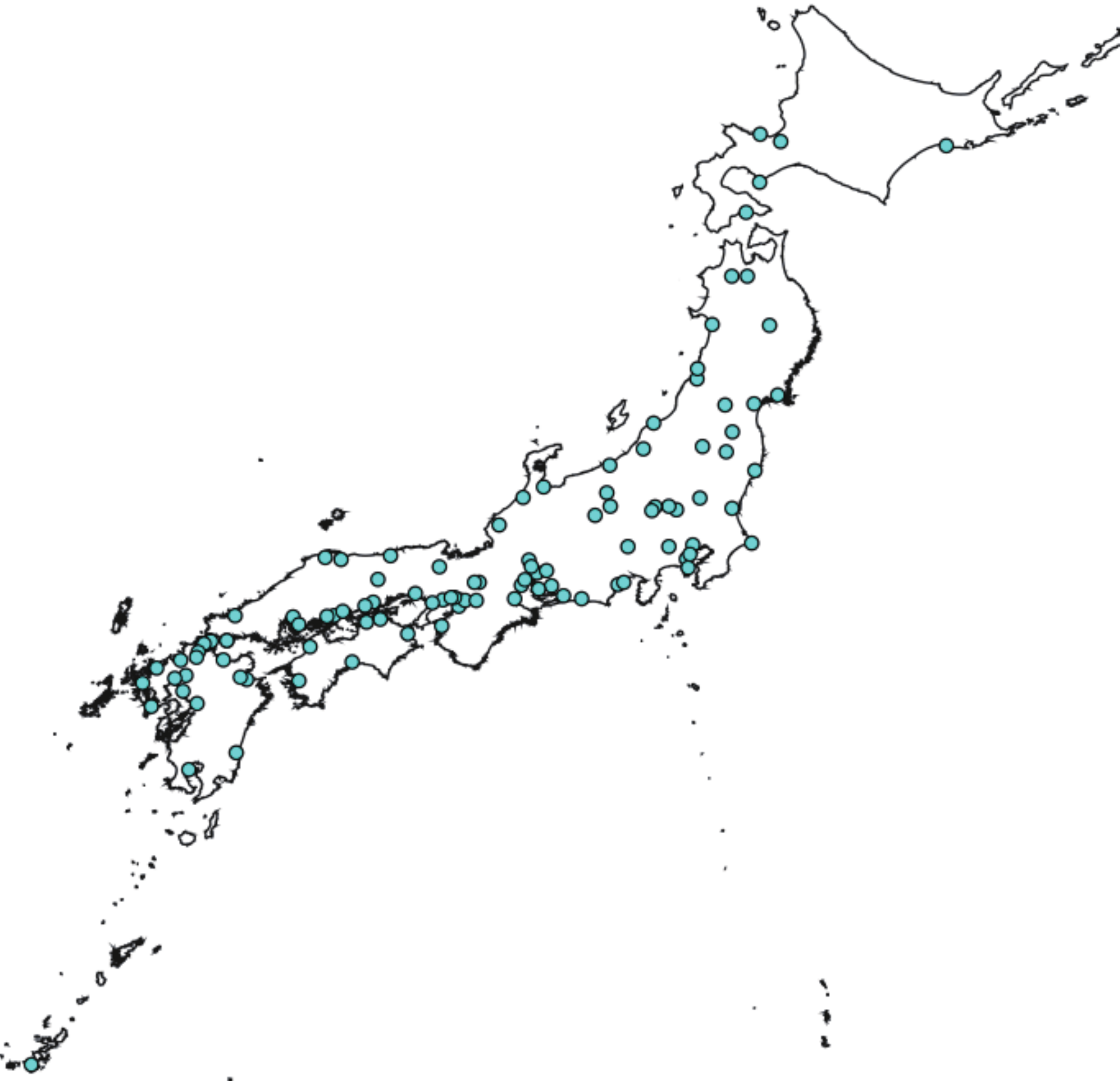}}
    \subfloat[26 large cities]{\label{fig:map26}\includegraphics[width=0.35\textwidth]{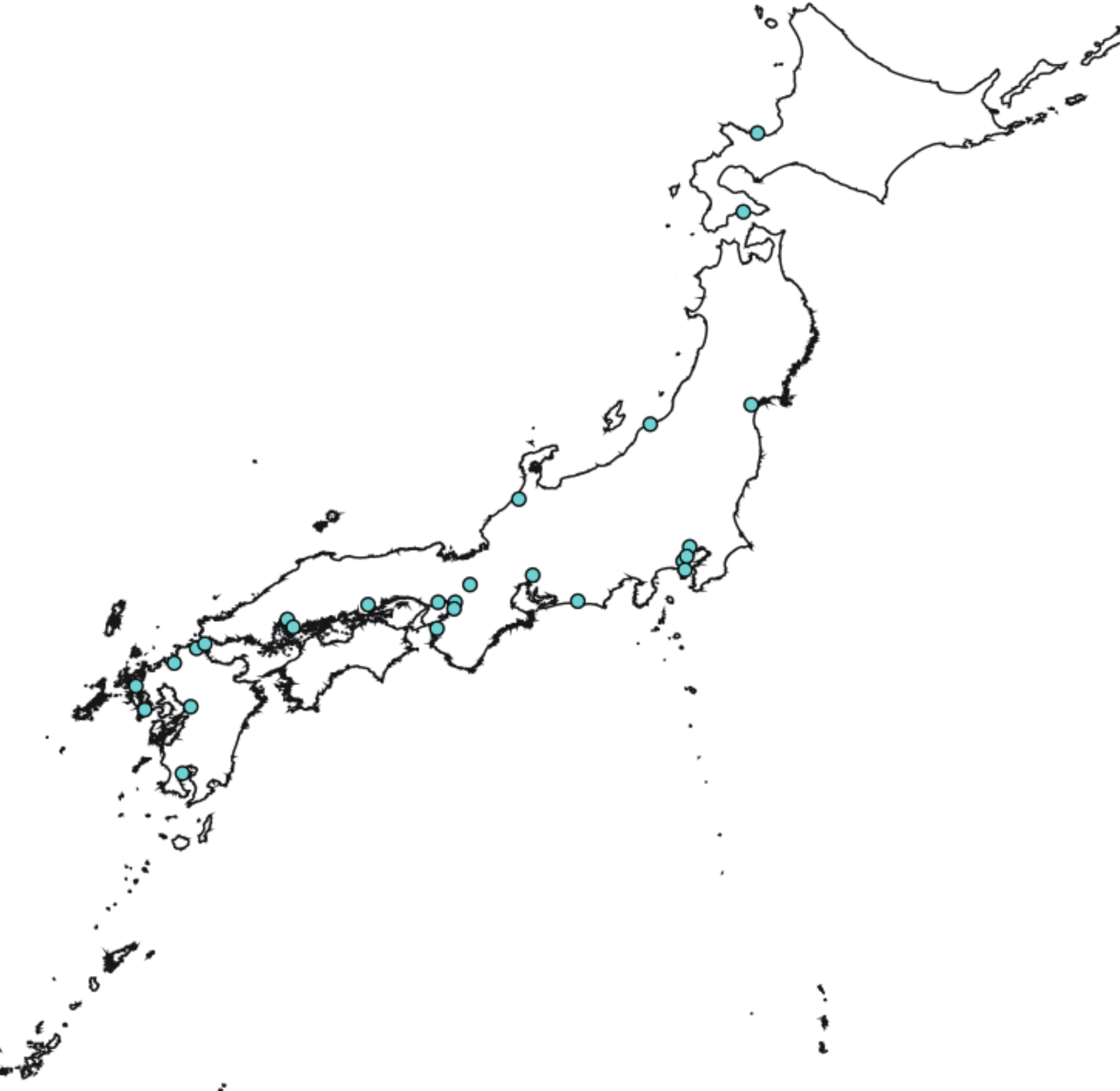}}
    \caption{Spatial distribution of the cities}
    \label{fig:map}
    \scriptsize{\begin{minipage}{400pt}
    Notes:
	Each circle indicates the location of the cities based on latitude and longitude information.
	Source:	Data on latitude and longitude are from the database of the Geospatial Information Authority of Japan.
    \end{minipage}}
\end{figure}

Fig.~\ref{fig:map} illustrates the spatial distribution of our sample cities.
All 26 large cities are included in the 108 cities.
The data on latitude and longitude are from the database of the Geospatial Information Authority of Japan: \url{http://www.gsi.go.jp/KOKUJYOHO/kenchokan.html}.
Shapefile was obtained from the website of the Geospatial Information Authority of Japan: \url{http://www.gsi.go.jp/kankyochiri/gm_jpn.html} (accessed 15 August 2018).

\clearpage
\section{Robustness check}\label{app:rob}
\setcounter{table}{0} \renewcommand{\thetable}{C.\arabic{table}}
\setcounter{figure}{0} \renewcommand{\thefigure}{C.\arabic{figure}}

\subsection{Placebo test using lead values of the typhoid death rate} \label{rob:lead}

\begin{table}[!h]

\begin{center}
\caption{Placebo test using lead values of the typhoid death rate}
\label{tab:lead}
\scriptsize

\begin{tabular}{lcccccc}
\toprule
&\multicolumn{3}{c}{Non-typhoid death rate}
&\multicolumn{3}{c}{MRE death rate}\\
\cmidrule(rrl){2-4}
\cmidrule(rrl){5-7}

&(1)&(2)&(3)&(4)&(5)&(6)\\
\hline

\textit{Typhoid}
&1.167***&1.179***&1.077** &0.743**&0.649*&0.562\\
&(0.368)&(0.401)&(0.417)&(0.311)&(0.378)&(0.384)\\
\textit{Typhoid}$_{t+1}$
&0.111&0.245&0.237&0.051&0.033&0.008\\
&(0.268)&(0.323)&(0.361)&(0.203)&(0.206)&(0.222)\\
\textit{Typhoid}$_{t+2}$
&&-0.073&-0.362&&-0.222&-0.150\\
&&(0.338)&(0.421)&&(0.402)&(0.368)\\
\textit{Typhoid}$_{t+3}$
&&&-0.252&&&0.058\\
&&&(0.504)&&&(0.315)\\

\hline

All control variables
&Yes&Yes&Yes&Yes&Yes&Yes\\
City- and Year- fixed effects
&Yes&Yes&Yes&Yes&Yes&Yes\\
Linear time trends
&Yes&Yes&Yes&Yes&Yes&Yes\\


Number of clusters
&107&106&101&28&26&26\\
Number of observations
&1,287&1,131&989&260&250&241\\

\bottomrule
\end{tabular}

{\scriptsize
\begin{minipage}{350pt}
\textit{Notes}: 
The non-typhoid death rate is defied as the number of deaths from all cause other than typhoid fever per 1,000 people.
The MRE death rate is defined as the number of the deaths from tuberculosis, pneumonia, bronchitis, meningitis, and heart disease per 1,000 people.
The number of clusters in column (4) is larger than that in Table~\ref{tab:cause} because we excluded Sapporo and Shizuoka city that did not have modern waterworks during our study period from the estimates reported in Table~\ref{tab:cause}.
***, **, and * represent statistical significance at the 1\%, 5\%, and 10\% levels, respectively.
Cluster-robust standard errors are in parentheses. 
\end{minipage}
}

\end{center}
\end{table}


We include the leads of the typhoid death rate in years $t+1$, $t+2$, and $t+3$ in the baseline specification reported in column (3) in Table~\ref{tab:non} as the placebo experiments.
If the coefficients on the leading variables were estimated to be significantly positive, the common trend assumption in the fixed-effects model might be violated and/or our baseline results would be affected by the omitted variables because the lead values are the realization values in the future periods.

The estimates are reported in Table~\ref{tab:lead}.
Columns (1)--(3) and (4)--(6) report estimates for the non-typhoid death rate and MRE death rate, respectively.
The demographic, socioeconomic, and meteorological controls as well as fixed effects are included in all specifications.
The estimated coefficients on the lead variables do not differ significantly from zero and those impacts are very weak in all specifications.
In contrast to the lead variables, the estimated coefficients on the typhoid death rate at year $t$ remain statistically significantly different from zero in almost all specifications.
This result implies both that the parallel-trend assumption should hold and that the omitted variable bias may not confound our baseline estimates reported in Table~\ref{tab:non} and \ref{tab:cause}.

\subsection{Effects of other waterborne death on the MRE death rate} \label{tab:ivfeoh}

\begin{table}[t]
\captionsetup{justification=centering,margin=1cm}
\begin{center}
\caption{Effects of other waterborne death on the MRE death rate, fixed effects two-stage least squares}
\label{tab:ivfeoh}
\scriptsize

\begin{tabular*}{135mm}{@{\extracolsep{\fill}}lccccc}
\toprule
Panel A: Second stage results
&&&&\\
&\multicolumn{5}{c}{Independent variable}\\
\cmidrule(rrrrl){2-6}
&Paratyphoid&Dysentery&Cholera&Diarrhea&All\\
\hline

Estimated coefficients on \textit{MRE death}
&476.179&23.756&112.676&$-$19.279&$-$111.797\\
&(1237.827)&(26.385)&(129.520)&(62.214)&(2131.402)\\

\hline

Control variables
&Yes&Yes&Yes&Yes&Yes\\

City- and Year- Fixed effects
&Yes&Yes&Yes&Yes&Yes\\
Linear time trends
&Yes&Yes&Yes&Yes&Yes\\

Number of clusters
&26&26&9&26&26\\
Number of observations
&258&258&116&258&258\\

\hline

\\
Panel B: First stage results
&&\\
&\multicolumn{5}{c}{Dependent variable}\\
\cmidrule(rrrrl){2-6}
&Paratyphoid&Dysentery&Cholera&Diarrhea&All\\
\hline

\textit{Water}
&$-$0.0000&$-$0.0010&0.0001&0.0012&0.0002\\
&(0.0001)&(0.0012)&(0.0001)&(0.0043)&(0.0046)\\
\hline

All control variables
&Yes&Yes&Yes&Yes&Yes\\
City- and Year- fixed effects
&Yes&Yes&Yes&Yes&Yes\\
Linear time trends
&Yes&Yes&Yes&Yes&Yes\\
Kleibergen-Paap \textit{rk} Wald $F$-statistic
&0.11&0.64&0.90&0.08&0.00\\
$R$-squared
&0.602&0.886&0.586&0.898&0.852\\
Number of clusters
&26&26&9&26&26\\
Number of observations
&258&258&116&258&258\\

\bottomrule
\end{tabular*}

{\scriptsize
\begin{minipage}{130mm}
\textit{Notes}: 
Observations are at the city-year level.
The MRE death is defined as the death from tuberculosis, pneumonia, bronchitis, meningitis, and heart disease.
The variables paratyphoid, dysentery, cholera, and diarrhea are the number of deaths from each disease per 1,000 people.
***, **, and * represent statistical significance at the 1\%, 5\%, and 10\% levels, respectively.
Cluster-robust standard errors are in parentheses. 
\end{minipage}
}

\end{center}
\end{table}

Although we assumed that typhoid fever was the key factor contributing to the Mills--Reincke phenomenon, there was a possibility that other waterborne diseases were also the factors of the phenomenon.
To test this hypothesis, we investigate the effects of paratyphoid, dysentery, cholera, and diarrhea death rate on the MRE death rate employing FE-2SLS model.
As well as typhoid fever, these diseases are waterborne and thus were likely to decrease due to the popularization of modern water-supply systems.
If the estimated coefficient of these death rates in second stage is positive and statistically significant, this result suggests that improvements of water quality affected the MRE deaths by preventing not only typhoid but also other waterborne diseases.
In this case, the exclusion restriction is broken.

Panel A of Table~\ref{tab:ivfeoh} reports the results of the FE-2SLS estimation using paratyphoid, dysentery, cholera, diarrhea, and all the above instead of the typhoid death rate.
We found no significant relationships between the MRE death rate and these dependent variables.
This result suggests that typhoid fever is the unique disease leading to the Mills--Reincke phenomenon and supports the assumption that modern waterworks affected the MRE deaths only through its effect on typhoid fever.
However, the first stage results in Panel B of Table~\ref{tab:ivfeoh} indicate that the instrument is irrelevant to waterborne death variables.
Modern waterworks might not decrease deaths due to waterborne diseases other than typhoid fever, at least directly.

\subsection{Flexible estimates}

\begin{table}[!b]

\begin{center}
\caption{Flexible estimates: effects of typhoid fever on non-typhoid and MRE deaths}
\label{tab:fle}
\scriptsize

\begin{tabular*}{160mm}{lcccccccc}
\toprule

&\multicolumn{4}{c}{Non-typhoid death rate}&\multicolumn{4}{c}{MRE death rate}\\
\cmidrule(rrrl){2-5}
\cmidrule(rrrl){6-9}
&(1)&(2)&(3)&(4)&(5)&(6)&(7)&(8)\\
\hline

\textit{Typhoid}
&0.943***&1.258**&1.324**&1.557***&0.467**&0.960***&0.829**&1.721*\\
&(0.315)&(0.460)&(0.510)&(0.507)&(0.223)&(0.217)&(0.421)&(0.977)\\

\hspace{6truept}$\times$\textit{After 1930}
&&$-$0.580&&&&$-$0.754***&&\\
&&(0.590)&&&&(0.279)&&\\

\hspace{6truept}$\times$\textit{West}
&&&$-$0.436&&&&$-$0.101&\\
&&&(0.620)&&&&(0.452)&\\

\textit{Water} in first stage
&&&&&&&&$-$0.008***\\
&&&&&&&&(0.003)\\

\hline

All Control variables
&Yes&Yes&Yes&Yes&Yes&Yes&Yes&Yes\\

City- and Year- fixed effects
&Yes&Yes&Yes&Yes&Yes&Yes&Yes&Yes\\

Linear time trends
&No&Yes&Yes&Yes&No&Yes&Yes&No\\


Number of clusters
&108&108&108&26&26&26&26&26\\
Number of observations
&1,457&1,457&1,457&258&258&258&258&258\\

\bottomrule\end{tabular*}

{\scriptsize
\begin{minipage}{155mm}
Notes: 
Observations are at the city-year level.
The non-typhoid death rate is defied as the number of deaths from all cause other than typhoid fever per 1,000 people.
The MRE death rate is defined as the number of the deaths from tuberculosis, pneumonia, bronchitis, meningitis, and heart disease per 1,000 people.
All regressions include the indicator variable for the Great Kant\=o Earthquake.
Details of the data sources of each variable used in the regression are provided in the text and the \ref{app:data}.
***, **, and * represent statistical significance at the 1\%, 5\%, and 10\% levels, respectively.
Cluster-robust standard errors are in parentheses.
\end{minipage}
}

\end{center}
\end{table}

We also conduct regression analyses using different specifications to explore the heterogeneity of the effect of typhoid fever on the death rates of other causes and confirm the robustness of our estimates.
Table \ref{tab:fle} shows the results.
Columns (1)--(4) and (5)--(8) report the estimates for the non-typhoid death rate and MRE death rate, respectively.
Column (8) presents the results of the FE-2SLS model and the other columns present the results of the FE model.
The estimated coefficients of \textit{Typhoid} in columns (1), (5), and (8), where the specifications do not include the city-specific linear time trends, are positive and statistically significant following the main results shown in Table \ref{tab:non} and Table \ref{tab:cause}.
These robust results support our hypothesis that typhoid fever is the key factor behind the Mills--Reincke phenomenon.
We create an indicator variable taking one if year \textit{t} is after 1930 and include \textit{Typhoid} $\times$ \textit{After 1930} in the specifications in columns (2) and (6).
The estimated coefficient of the interaction term is insignificant for the non-typhoid death rate but significantly negative for the MRE death rate.
This result implies that severe typhoid fever, which tends to cause complications, decreased more in the 1920s than in the 1930s.
In columns (3) and (7), we also investigate the heterogeneity of the effect of \textit{Typhoid} across areas using the dummy variable \textit{West} taking one if city \textit{i} is in western Japan.
The results indicate no significant difference in the effect between eastern and western Japan; that is, a decrease in typhoid fever contributed to a decrease in deaths due to other causes throughout the country.
Column (4) shows the results estimated from the same sample following the estimates in Table 4.
The estimated coefficient of the typhoid death rate is positive and significant, highlighting that the result for the non-typhoid death rate is stable regardless of the number of observations.

\setcounter{equation}{0} \renewcommand{\theequation}{\arabic{equation}}
\section{Cost-benefit analysis} \label{app:cba}
It is important to investigate not only benefits but also costs to debate about the efficacy of the public health improvements in developing countries (\eg, Cutler and Miller 2005; Ferrie and Troesken 2008).
Taking advantage of our rich data, we estimate the benefits of chain effects of modern water-supply systems by cause of death and costs of installation in Japan between 1922 and 1936.
This contributes to understanding a comprehensive and detailed economic advantage provided by improvements in water-supply systems. 
In our analysis, we focus on the benefits that could be gained from decline in deaths from other causes.
However, a decline in typhoid fever by using clean water were likely to bring about many kinds of benefits (\eg, Beach \etal, 2016).

To calculate average annual cost between 1922 and 1936, we compile the information both on the total cost of each laying and expansion project and on the exact date of their completion in each city from the SRWS and SWS.
Following a previous study, we assume that the facilities of the modern water-supply systems could be used for only 10 years from completion (see Cutler and Miller 2005, p.18; Beach \etal~2016, p.71).%
\footnote{This assumption is quite conservative because the average lifetime of the water-supply system in postwar Japan is considered as approximately 50 years (see Japan Water Works Association 1967a, p.505).}
The cost are calculated according to the following equation.
\begin{align}
\text{\textit{Cost per year}}_{p}&= \frac{\text{\textit{Total cost}}_{p}}{10}\\
\text{\textit{Cost}}_{p} &=
\begin{cases}
\text{\textit{Cost per year}}_{p} \times \left(\text{\textit{Year}}_{p} + 10 - 1921\right) &\text{if}\ 1912 \le \text{\textit{Year}}_{p} < 1922\\
\text{\textit{Cost per year}}_{p} \times 10 &\text{if}\ 1922 \le \text{\textit{Year}}_{p} \le 1926\\
\text{\textit{Cost per year}}_{p} \times (1936 - \text{\textit{Year}}_{p}) &\text{if}\ 1926 < \text{\textit{Year}}_{p} < 1936\\
0 &\text{else}
\end{cases}
\\
\text{\textit{Cost}}&= \frac{\sum_{p}\text{\textit{Cost}}_{p}}{15}
\end{align}
where \textit{p} indexes projects of modern waterworks installation.
The variable \textit{Total cost}$_{p}$ is the total cost of the project \textit{p}, \textit{Year}$_{p}$ is the year of the completion, and \textit{Cost} is the average total cost per year between 1922 and 1936.
Accordingly, the average annual cost in 2014 dollars is estimated as 113 million dollars.%
\footnote{We convert the costs in 1930 yen into costs in 2014 dollars using the CPI-U of the United States (16.7 in 1930) and the rate of exchange.
The data on the rate of exchange in 1930 between Japanese yen and the US dollar (100 yen compared to 49.375 dollars) are taken from Financial Bureau of the Ministry of Finance (1931).}

The benefits of chain effects of modern waterworks are calculated as follows.
First, we estimate the number of people saved from cause-specific deaths due to clean water via decline in typhoid deaths from 1922 to 1936.
\begin{align}
\text{\textit{Water}}_{t} &= \frac{\sum_{i}tap_{it}}{\sum_{i}pop_{it}} \times 100\\
\text{\textit{Saved people}}_{\text{\textit{Typhoid}, }t} &= \text{\textit{Water}}_{t} \times \delta_{\text{\textit{Typhoid}}} \times \frac{\sum_{i}pop_{it}}{10000}\\
\text{\textit{Saved people}}_{\text{\textit{Typhoid}}} &= \frac{\sum_{t}\left(\text{\textit{Saved people}}_{\text{\textit{Typhoid}, }t}\right)}{15}
\end{align}
where \textit{tap}$_{it}$ is the number of water taps and \textit{pop}$_{it}$ is the population in city \textit{i} in year \textit{t}.
$\delta_{\textit Typhoid} (= 0.0080)$ is the estimated coefficient of the coverage of tap water on typhoid death rate reported in column (1) in Panel B of Table~\ref{tab:cause}.
Using this point estimate, the coverage of tap water in year \textit{t} (\textit{Water}$_t$), and the population size in year \textit{t} ($\sum_{i}pop_{it}$), we yield the number of people saved from typhoid deaths due to clean water (\textit{Saved people$_{\text{Typhoid}}$}).
In this calculation, we assume that the effect of the popularization of tap water did not vary by either population size or the initial tap water coverage rates.
And then, we estimate the number of people saved by chain effects for each disease:
\begin{align}
\text{\textit{Saved people}}_{d} &= \text{\textit{Saved people}}_{\text{\textit{Typhoid}}} \times \delta_{d}
\end{align}
where $\delta_{\textit{d}}$ is our estimated effect of typhoid fever on the death from cause-specific disease \textit{d}.

Second, we calculate the average life expectancy of people who would have died if they had not received any benefit from the waterworks.
Therefore, we compile data on the number of cause-specific deaths by sex and age groups in Japan between 1922 and 1936 and on the life expectancy by sex and age groups in 1930.%
\footnote{The data on life expectancy are taken from the Statistics Bureau of the Cabinet (1936).}
Our calculus equation is given by
\begin{align}
\text{\textit{Death age}}_{d} &= \frac{\sum_{\textit{sex}, \textit{age}}\left(\textit{Death}_{\textit{sex}, \textit{age}, d} \times \text{\textit{age}}\right)}{\sum_{\textit{sex}, \textit{age}}\textit{Death}_{\textit{sex}, \textit{age}, d}}
\end{align}
where \textit{Death}$_{sex. age, d}$ is the number of deaths from the disease \textit{d} by sex \textit{sex} and age groups \textit{age}, and \textit{Death age}$_{d}$ means the average age of death from \textit{d}.
Hence, the average life expectancy who died from disease \textit{d} (\textit{Life expectancy}$_d$) equals the life expectancy at \textit{Death age}$_{d}$.

Third, we estimate the value of a person-year at that time.
Following Cutler and Miller (2005), we assume that a reasonable value of a person-year in 2003 was about 100,000 dollars on average (Viscusi and Aldy 2003) and, for simplicity, that the value of a person does not vary with age.
Under this assumption, in 2014 dollars, the average value of a person-year in 2003 is 128,660 dollars.%
\footnote{The consumer price index for all urban consumers (CPI-U) was 184 in 2003 and 236.736 in 2014, respectively. The CPI-U data are publicly available at the website of the Bureau of Labor Statistics, United States Department of Labor (\url{http://www.bls.gov/cpi/}). We use Table 24 of the CPI Detailed Report Data for September 2015 for our calculation.}
In addition, we assume that the elasticity of the value of life with respect to the per capita gross national product to range from 1.5 to 1.7 (Costa and Kahn 2002).
Since the gross domestic product per capita was 1,850 Geary--Khamis dollars in Japan in 1930 and 29,459 Geary--Khamis dollars in the United States in 2003,
the value of a person-year in Japan in 1930 is conservatively estimated as 4,753 in 2014 dollars.%
\footnote{The data on the gross domestic product per capita in 1990 Geary--Khamis dollars are obtained from the database of The Maddison Project, \url{http://www.ggdc.net/maddison/maddison-project/home.htm}, 2013 version.}

Finally, the average benefits of chain effects of modern waterworks for each disease \textit{d} (\textit{Benefit}$_d$) are calculated as follows:
\begin{align}
\text{\textit{Benefit}}_{d} &= \textit{Saved people}_{d} \times \textit{Life expectancy}_{d} \times \textit{Value of a person-year} \nonumber\\
& = \textit{Saved person-years}_{d}  \times 4753
\end{align}
%

\setcounter{table}{0} \renewcommand{\thetable}{D.\arabic{table}}
\begin{table}[!h]

\begin{center}
\caption{Cost--benefit analysis of chain effects of modern water-supply systems}
\label{tab:srr}
\scriptsize

\begin{tabular}{lrrrrrr}
\toprule
&\multicolumn{3}{c}{Fixed effect}
&\multicolumn{3}{c}{FE-2SLS}\\
\cmidrule(rrl){2-4}
\cmidrule(rrl){5-7}
&Benefits&\multicolumn{1}{l}{Social rate}&\multicolumn{1}{l}{Ratio to}&Benefit&\multicolumn{1}{l}{Social rate}&\multicolumn{1}{l}{Ratio to}\\
&&of return (\%)&typhoid (\%)&&of return (\%)&typhoid (\%)\\
\hline
\textit{Non-typhoid}
&142&126.2&97.4&-&-&-\\
\textit{MRE death}
&117&103.6&80.0&463&410.9&317.3\\
Tuberculosis
&(17)&(14.8)&(11.4)&103&90.9&70.2\\
Pneumonia
&79&69.9&54.0&209&185.8&143.5\\
Bronchitis
&(9)&(7.8)&(6.0)&37&32.8&25.3\\
Meningitis
&(12)&(10.9)&(8.4)&133&118.2&91.3\\
Heart disease
&(4)&(3.3)&(2.5)&(5)&(4.9)&(3.8)\\

\bottomrule
\end{tabular}

{\scriptsize
\begin{minipage}{330pt}
\textit{Notes}:
Benefits is in 2014 dollars in millions.
The insignificant results are in parentheses.
\end{minipage}
}

\end{center}
\end{table}

Table~\ref{tab:srr} reports the results of our cost--benefit analysis.
Social rate of return in third and sixth columns is defined as benefit divided by cost.
Ratio to typhoid in fourth and seventh columns is calculated as the ratio of benefit of decline in each cause-specific death to benefit of decline in typhoid fever due to modern water-supply systems.
The results that are insignificant in our estimates are in parentheses.
Surprisingly, the benefit alone of chain effect for MRE deaths (117 and 463) exceeded the average total cost regardless of identification strategy.
In particular, the result of FE-2SLS exhibits that social rate of return was 410.9\%, that is, the benefit was approximately 4.1 times more than the cost.
Moreover, the benefit of decline in MRE deaths was about three times than the benefit of decline in typhoid deaths itself.
This result suggests that improvements of water-supply systems provided economic advantage even more than previously expected through the reduction in non-waterborne deaths.

\renewcommand{\refname}{References used in the Appendices}

\renewcommand{\refname}{Statistical reports}

\end{document}